\documentclass[aps, prl, twocolumn, groupedaddress, superscriptaddress, shortbibliography, notitlepage]{revtex4-1}

\usepackage{graphicx} 
\usepackage{color} 
\usepackage{amsmath}
\usepackage{amsfonts} 
\usepackage{amssymb} 
\usepackage{dcolumn}
\usepackage[justification=centerlast]{caption}
\usepackage{enumitem}
\usepackage{xr}
\usepackage{hyperref}
\hypersetup{colorlinks=true, urlcolor=blue,
  linkcolor=blue, citecolor=blue}
\usepackage{subcaption}

\begin{document}
\title{Quantifying hidden order out of equilibrium}
\author{Stefano Martiniani}
\email{smartiniani@nyu.edu}
\affiliation{Center for Soft Matter Research, Department of Physics, New York University, New York 10003, USA}
\author{Paul M. Chaikin} 
\email{chaikin@nyu.edu}
\affiliation{Center for Soft Matter Research, Department of Physics, New York University, New York 10003, USA}
\author{Dov Levine} 
\email{dovlevine19@gmail.com}
\affiliation{Department of Physics, Technion - IIT, 32000 Haifa, Israel}

\begin{abstract}
While the equilibrium properties, states, and phase transitions of interacting systems are well described by statistical mechanics, the lack of suitable state parameters has hindered the understanding of non-equilibrium phenomena in diverse settings, from glasses to driven systems to biology. The length of a losslessly compressed data file is a direct measure of its information content: The more ordered the data is, the lower its information content and the shorter the length of its encoding can be made. Here, we describe how data compression enables the quantification of order in non-equilibrium and equilibrium many-body systems, both discrete and continuous, even when the underlying form of order is unknown. We consider absorbing state models on and off-lattice, as well as a system of active Brownian particles undergoing motility-induced phase separation. The technique reliably identifies non-equilibrium phase transitions, determines their character, quantitatively predicts certain critical exponents without prior knowledge of the order parameters, and reveals previously unknown ordering phenomena. This technique should provide a quantitative measure of organization in condensed matter and other systems exhibiting collective phase transitions in and out of equilibrium.
\end{abstract}
\maketitle
Intuitively, the more ordered a system is, the shorter the description required to specify a typical microstate.  If the probability distribution of the ensemble of microstates is known, then the Shannon entropy \cite{shannon2001mathematical} provides a quantitative measure of the information content and order. For a random variable $X$ the Shannon entropy is defined as
\begin{equation}
H(X) = -\sum_{\{x\}} p(x) \log p(x),
\label{eq-entropy}
\end{equation}
which may be thought of as the average uncertainty in $X$.  Here, $p(x)$ is the probability that a given signal $x$ is generated by a given source; in physics terms, this may be thought of as defining an ensemble.  If we take $x$ to specify microstates of a thermodynamic ensemble, and $p(x)$ to be the probabilities of their occurrence, then Eq.~\ref{eq-entropy} reproduces the thermodynamic entropy appropriate to this ensemble.  It is important to understand that the framework of equilibrium statistical thermodynamics provides \textit{a-priori} probabilities, but this is not the case for systems out of equilibrium, making the explicit computation of $H$ in general impossible \footnote{$H$ may be estimated from block entropies (based on the occurrence frequencies of blocks of finite size). Sampling issues attendant to this method are discussed in \cite{grassberger2003entropy}.}\nocite{grassberger2003entropy}.  

Knowledge of the probability distribution is not required for the algorithmic approach to information content pioneered by Kolmogorov and Chaitin \cite{kolmogorov1968three,chaitin1966length}.  This approach culminated in the definition of the \textit{Kolmogorov complexity} $K$ \cite{cover2012elements}, as (loosely speaking) the length of the shortest computer program able to generate a given data sequence.  Under fairly general conditions, $H$ and $K$ are closely related, and become equal in the large system limit \cite{cover2012elements}.  However, although elegant, the Kolmogorov complexity is not typically computable, and so can not be used for physical systems. In this paper we study an easily accessible proxy for these measures, which we will refer to as \textit{Computable Information Density} (CID), which is proportional to the length of a losslessly compressed data string \footnote{Note that the CID is not the same as the compression ratio (or compressibility) $\varrho$ of the sequence, in fact $\text{CID}=\varrho \log_2 |\alpha|$, where $|\alpha|$ is the dictionary size of the sequence \cite{ziv1978compression}; see the SI for further discussion.}\nocite{ziv1978compression}.
Concretely we define:
\begin{equation}
\text{CID} \equiv \frac{\mathcal{L}(x)}{L}.
\end{equation}
where $\mathcal{L}(x)$ is the total binary code length of the compressed sequence, and $L$ is the length of the original sequence $x$ (the number of sites in the system).  We have used the LZ77 compression algorithm \cite{ziv1977universal} (though other choices of universal codes are available), with extrapolation to the thermodynamic limit performed according to Eq.~S11; see SI for a description of the algorithm and a discussion of the extrapolation.

The problem of finding a faithful minimum encoding is the province of lossless data compression, and is commercially important in data storage and telecommunication.  Shannon's source coding theorem \cite{shannon2001mathematical} states that (in the large system limit) the length of the shortest encoding a file can have without loss of information is $H$.  Thus, asymptotically optimal data compression algorithms may be used to approximate $H$ for a broad class of data\footnote{In particular, data generated by any stationary and ergodic process.  By stationary we mean that the probability of an event is invariant with time and by ergodic we mean that  the strong law of large numbers holds \textit{viz.} the sample average tends to the expected value $\lim_{n \to \infty}1/n \sum_{i=1}^n X_i = \mu$}, approaching it in the thermodynamic limit of large systems.  Unlike the Shannon (or ordinary thermodynamic) entropy, CID is well-defined for any given sequence \cite{merhav2015sequence}.  This has the important consequence that CID may be defined for finite portions of a system, allowing us to study its behavior and correlations in both space and time.

Data compression was first applied to the two-dimensional Ising model by Sheinwald, Lempel and Ziv \cite{sheinwald1990two} as a benchmark for image compression. More recently, application to statistical physics has been mostly through the analysis of the time dependence of single-site variables. For equilibrium systems, a time series of the spin or the Edwards-Anderson autocorrelation parameter at a given site, obtained by Monte Carlo simulation, was used to locate the critical points of the 3D Edwards-Anderson spin glass \cite{cortez2014phase} and the 2D and 3D Ising models \cite{vogel2012data, melchert2015analysis}, and to approximate the entropy of the 2D Ising model \cite{melchert2015analysis}.  Data compression has also proven a useful tool in the definition and characterization of complexity of one-dimensional dynamical models, such as cellular automata and  dynamical systems \cite{kaspar1987easily, steeb1997exact, benci2002dynamical, baronchelli2005measuring, grassberger2012randomness, aaronson2014quantifying, estevez2015lempel}, as well as for turbulence \cite{cerbus2013information}. Methods based on data compression have also been used to estimate the entropy production of a non-equilibrium stationary state \cite{roldan2010estimating, roldan2012entropy} and to detect the onset of chaos in biological systems \cite{nykter2008gene, galas2010biological, flann2013kolmogorov}.

In this paper, we study the extent to which we can operationally define and use data compression in many-body non-equilibrium systems, in particular those where the nature of ordering is unclear.  To this end, we study the compression of entire microstates, rather than time series of single variables. We consider several different interacting non-equilibrium systems, both on and off-lattice, in one and two dimensions \footnote{We have studied systems in three and four dimensions, with similar results.}.  We show that  CID provides an easily applied and quantitatively accurate measure of information content which can serve as a simple and sensitive way to quantify order, its evolution in time, and its dependence on control parameters \footnote{Different interesting measures, often termed ``complexity'', which are minimal in perfectly ordered and perfectly random cases, have also been proposed; see, \textit{e.g.} \cite{grassberger1986toward}.}\nocite{grassberger1986toward}. In particular, we show that non-equilibrium analogs of both discontinuous and continuous phase transitions are well characterized by singularities in CID, that certain critical exponents can be extracted without \textit{a priori} knowledge of the order parameter, and that previously unknown ordering phenomena can be discovered.

\begin{figure*}
\includegraphics[width=\linewidth]{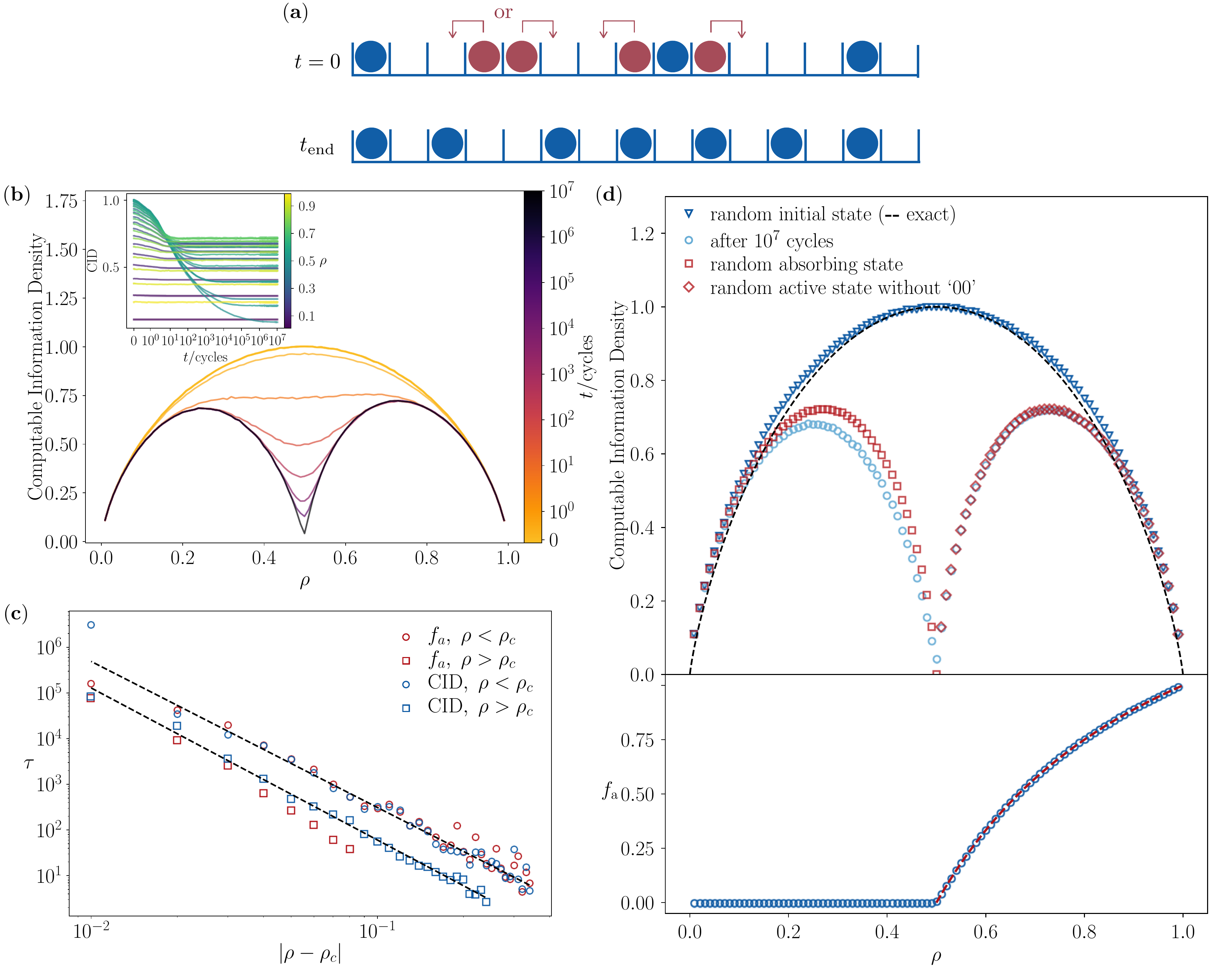}
\protect\caption{The 1D Conserved Lattice Gas model of size $L=10^5$.  (a) At time $t=0$ the system is in an active randomly sampled  state (active sites in red) and the possible moves prescribed by the dynamics are indicated by the arrows. When the particle density is below the critical density $\rho_c$, the system relaxes to an absorbing state, such that the fraction of active sites $f_a=0$. (b) Time dependence of the CID as a function of particle density $\rho$, a cycle corresponds to $L$ randomly attempted moves. The system orders as a function of time, developing a cusp-minimum at the critical density $\rho_c=0.5$. The inset shows the CID time evolution profile for several densities. (c) Characteristic time $\tau$ as a function of $|\rho - \rho_{c}|$, as measured by the decay of $f_{a}$ and CID, showing  identical relaxation and critical exponent $\nu_\parallel = 3 \pm 0.3$ from both measures. Lines of best fit (dashed black lines) were obtained by bootstrapped minimum mean square error fits using a robust covariance estimator \cite{hubert2010minimum, efron1994introduction}. (d) Top panel: comparison of the random initial states, states found by the dynamics after $10^7$ cycles, uniformly sampled absorbing states below $\rho_c$, and active states without $00$ pairs above $\rho_c$. Lower panel: fraction of active sites as a function of $\rho$, red dashed line is the exact solution from Ref.~\cite{de2005conserved}. \label{fig:1}} 
\end{figure*}

To illustrate the use of CID, we consider a particularly simple model with a non-equilibrium phase transition, the Conserved Lattice Gas (CLG) in 1D.  Initially, $N$ particles are distributed randomly on $L \geq N$ sites with no multiple occupancy.  An occupied site is considered `active' if one of its neighbors is also occupied.  The dynamics consist of moving particles randomly from active sites to unoccupied neighboring sites, as illustrated  in Fig.~\ref{fig:1}a (in practice we implement random sequential updates, so we displace one particle at a time).  The statistical state of the system is characterized by the order parameter $f_a$, the fraction of sites that are active.  An `absorbing state' is attained when $f_a =0$, at which point the dynamics ends.  No absorbing states are possible for densities $\rho \equiv N/L$ higher than the geometrical limit $\rho_{G} = 0.5$.  For absorbing state models in general \cite{Non-Equilibrium_Book}, it is well known that there exists a critical density $\rho_{c}$, such that for $\rho>\rho_c$ the system evolves to an active, fluctuating steady state with a well-defined $f_a >0$, while for $\rho < \rho_c \leq \rho_{G}$, the system evolves to an absorbing state.  The 1D CLG is atypical in the sense that $\rho_{c} = \rho_{G}$ \cite{Non-Equilibrium_Book}, but will be seen to have non-trivial correlations in the absorbing phase.  For the 1D CLG the total number of possible absorbing state configurations is $\binom{(1-\rho)L}{\rho L}$ when $\rho\leq\rho_G$ and $0$ otherwise; an example of one such state is shown in Fig.~\ref{fig:1}a for $t=t_{\text{end}}$. States of the 1D CLG may be represented simply as a binary string of 0's and 1's signifying the occupation of the sites. These strings can easily be compressed by a large variety of universal codes, we do so by the `unrestricted' Lempel Ziv string-matching code, also known as LZ77 algorithm \cite{ziv1977universal}.

\begin{figure}
\includegraphics[width=\linewidth]{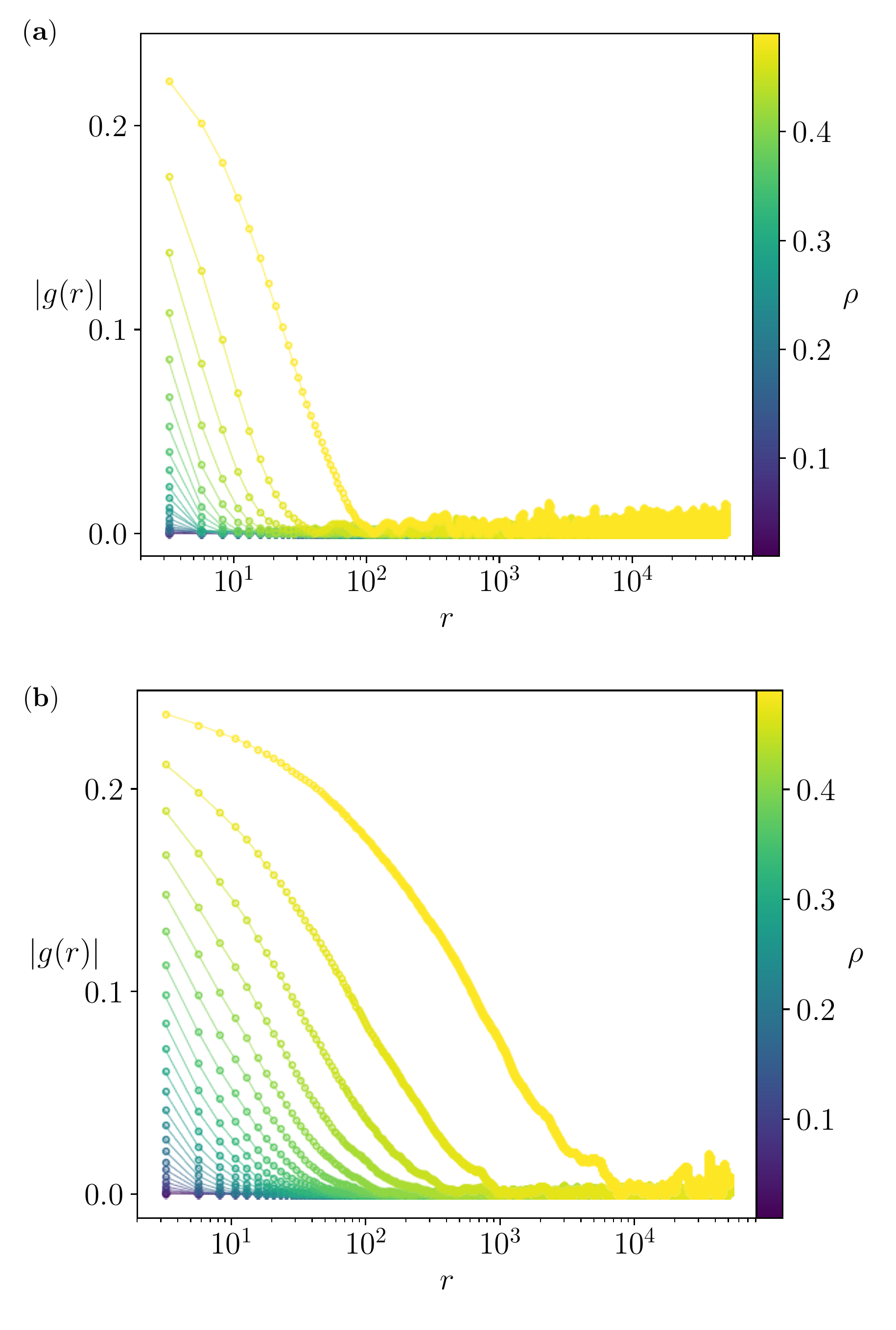}
\protect\caption{Absolute value of the autocorrelation function $g(r) \equiv \langle x_i x_{i+r} \rangle - \langle x \rangle^2$ for the 1D Conserved Lattice Gas model for (a) uniformily sampled absorbing states and (b) absorbing states arrived at by the dynamics. Clearly the 1D CLG dynamical states are correlated over a much larger scale. \label{fig:2}}
\end{figure}
 
We analyze a 1D CLG model of size $L=10^5$ with periodic boundary conditions, for $99$ densities in the range $0.01\leq \rho \leq 0.99$. Starting from random (Bernoulli distributed) initial configurations, we let the system evolve for $10^7$ full cycles (sweeps) by random sequential updates. At regular time intervals we measure the CID by LZ77. 

In Fig.~\ref{fig:1}b we show the CID as a function of $\rho$ for different times, with the inset indicating the CID time evolution profiles. At time $t=0$ the CID matches the Shannon entropy of a Bernoulli sequence $H=-\rho \log \rho - (1-\rho) \log (1-\rho)$, and at low and high densities $\rho$ the entropy remains unchanged in time. For densities near the critical point $\rho_c = 0.5$ the system organizes in time and the entropy tends to $0$ at the critical point, where only a single absorbing state is allowed. We fit the time dependent CID in the inset of Fig.~\ref{fig:1}b and the fraction of active sites $f_a(t)$ (not shown) with the functional form $y(t) = (y_0 - y_\infty) e^{-t/\tau}(t/t_0)^{-\delta} + y_\infty$, where $y_0$ and $y_\infty$ are the zero and infinite time limits, respectively, and $\delta$ and $t_0$ are fitted parameters that are roughly constant for all densities. In Fig.~\ref{fig:1}c we show the characteristic time $\tau$ as a function $|\rho - \rho_c |$. Analysis of the CID reveals a divergence of the correlation time (critical slowing down) in quantitative agreement with measurements performed on the time decay of $f_a$, which is the standard order parameter for the analysis of this model. The transition is thus continuous in nature and a fit of $\tau$ shows a power-law divergence of the form $\tau \sim |\rho - \rho_{c}|^{-\nu_{\parallel}}$, with $\nu_{\parallel} = 3 \pm 0.3$. 

Analysis of the CID immediately shows the extent that the dynamics orders the states. In the upper portion of Fig.~\ref{fig:1}d we show the CID of the final (absorbing or stationary active) states as obtained by the dynamics (blue circles).  On the active side, $\rho > \rho_{c}$, we compare dynamically obtained states (blue circles) with uniformly sampled unrestricted active states (blue triangles), as well as with uniformly sampled active states with no `$00$' pairs (red diamonds) since they are disallowed by the dynamics \footnote{The 1D CLG dynamics does not allow for the creation of '00' pairs where none existed, a fact already noted by de Olivera \cite{de2005conserved}.}\nocite{de2005conserved}.  The perfect match in CID between this latter set and those obtained from the dynamics indicates that this is precisely the ensemble sampled by the dynamics.  That the CID of the unrestricted active states is much higher than these clearly highlights the degree to which the dynamically accessible states are more ordered than the unrestricted active states. 

Ordering due to the dynamics is even more dramatic in the absorbing phase $\rho < \rho_c$. Comparison of the CID between uniformly sampled  absorbing states (red squares) and those arrived at by the dynamics (blue circles) shows that the dynamical states are more ordered than the random absorbing states, with the relative gap between the two growing as $\rho \to \rho_c$. This shows that the dynamics sample only a small subset of ordered states out of all the possible absorbing states. To understand the nature of the ordering we compute the autocorrelation function $g(r) \equiv \langle x_i x_{i+r} \rangle - \langle x \rangle^2$ \footnote{We computed the autocorrelation function by Wiener-Khinchin ${\langle x_i x_{i+r} \rangle = \mathcal{F}^{-1}\left( |\mathcal{F}(x)|^2\right)}$, that is the inverse Fourier transform of the power spectral density, both computed by discrete FFT.} for the random absorbing states (Fig.~\ref{fig:2}a) and the dynamically sampled absorbing states (Fig.~\ref{fig:2}b). In both cases the values of $g(r)$ alternate between positive and negative values due to the effective nearest neighbor repulsion, but as $\rho \to \rho_c$ it is apparent that the correlations are much longer ranged for the dynamically sampled absorbing states than for the random absorbing states. These longer ranged correlations indicate that the dynamics spreads out the particles in a very uniform way as the critical point is approached, a point which was not appreciated in this model before it was revealed by CID.

\begin{figure*}
\includegraphics[width=\linewidth]{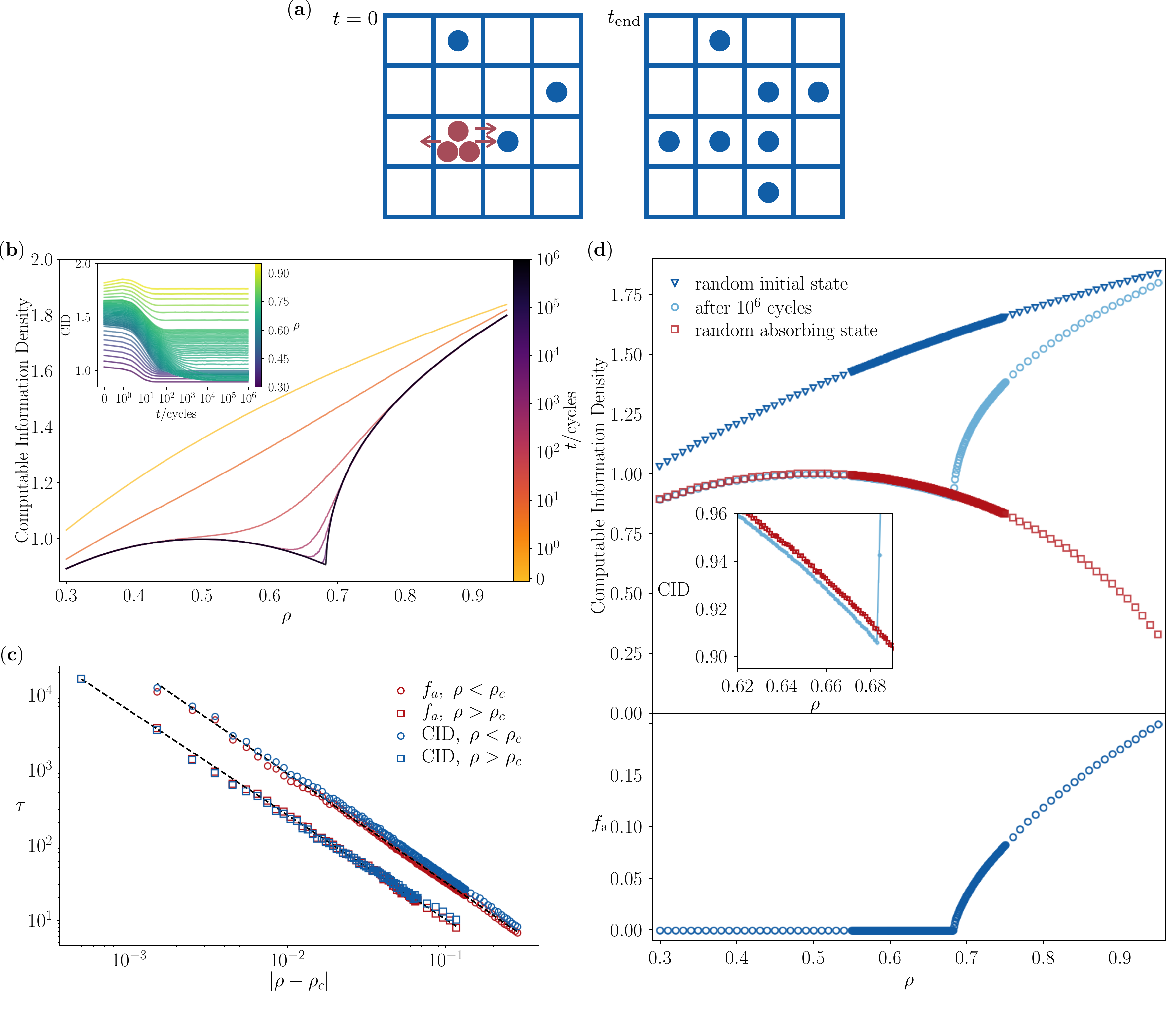}
\protect\caption{2D Manna model with size $2^{10} \times 2^{10}$ and $z_\text{max}=1$. (a) At time $t=0$ the system is in a randomly sampled state and sites occupied by more than $z_\text{max}$ particles are active (active sites in red). At each time step a randomly sampled active site is empitied by redistributing all particles to its nearest neighbors, a possible move is illustrated by the red arrows. When the particle density is below the critical density $\rho_c$, the system relaxes to an absorbing state, such that the fraction of active sites $f_a=0$. (b) CID as a function of particle density $\rho$ for different times, a cycle corresponds to $L$ randomly attempted moves. The system orders as a function of time, developing a cusp-minimum at the critical density $\rho_c \approx 0.683$. The inset shows the CID time evolution profile for several densities. (c) Characteristic time $\tau$ as a function of $|\rho - \rho_{c}|$, as measured by the decay of the fraction of active sites $f_{a}$ and the CID, showing  identical relaxation and critical exponent $\nu_\parallel = 1.3 \pm 0.2$ from both measures. Lines of best fit (dashed black lines) were obtained by bootstrapped minimum mean square error fits using a robust covariance estimator \cite{hubert2010minimum, efron1994introduction}. (d) Top panel: comparison of the random initial states, states found by the dynamics after $10^6$ cycles and uniformly sampled absorbing states; the inset shows that near the critical point the absorbing states arrived at by the dynamics have lower CID, and are more ordered than the uniformly sampled ones. Lower panel: fraction of active sites as a function of $\rho$. \label{fig:3}}
\end{figure*}

We next consider a two-dimensional system, a discrete lattice sandpile model known as the Manna model. In this model, the sites of a $M \times M$ square lattice (with periodic boundary conditions) are considered active if they are occupied by more than $z_{\text{max}}$ particles. The model allows for an unlimited number of particles at each site and the initial configuration is generated by depositing $N=\rho L$ particles at random on the lattice sites, where $L=M^2$ is the total number of sites. At each time step one active site is selected and all of its particles redistributed to the neighboring sites at random. This procedure is performed repeatedly, until either there are no active sites or the system arrives to a stationary (steady) state with a characteristic average fraction of active sites $f_a$.  Here, we take $z_{\text{max}}= 1$, such that $\rho_{G}=1$ and $\rho_c\approx 0.683$. An example of an initial random state and a final absorbing state is given in Fig.~\ref{fig:3}a.

In order to compute the CID of a two or higher dimensional system we flatten the grid.  In 2D, we use a Peano-Hilbert space filling curve \cite{hilbert1891ueber}; this is also known as a `Hilbert scan' and it requires $M=2^m$.  This scan, which covers the lattice in a self-similar fashion and preserves locality \footnote{In the sense that any two points close along the curve are also close in real space, though points nearby in space are not necessarily close on the curve.} has been shown to give optimal (distortion-free) compression as $L \to \infty$ \cite{lempel1986compression}.

In Fig.~\ref{fig:3} we show results for the 2D Manna model of size $2^{10} \times 2^{10}$ over $246$ densities in the range $0.3\leq\rho\leq0.95$, evolved for approximately $10^6$ full cycles. In the CLG the alphabet (possible site occupancies) is $\{0,1\}$, whereas in the Manna model the alphabet may contain any positive integer and its size may change as the system evolves. A reduction in alphabet size during the evolution contributes to a decrease in the CID. In Fig.~\ref{fig:3}b we show the time evolution of the CID as a function of $\rho$. The inset shows the CID time evolution profiles for some of the densities; the curves are averaged over $6$ independently sampled random initial conditions. Note that the CID is $> 1$ for the initial random configurations because the alphabet size is greater than 2. 

At long times, the system develops a sharp cusp-like minimum around $\rho_c \approx 0.683$  indicating a continuous phase transition, and the critical slowing-down is characterized in Fig.~\ref{fig:3}c. The correlation times measured from the CID and the fraction of active sites $f_a$ are in quantitative agreement with each other and we obtain critical exponent $\nu_{\parallel} = 1.3 \pm 0.2$, in agreement with the known value \cite{Non-Equilibrium_Book}. In Fig.~\ref{fig:3}c we compare the states after $10^6$ iterations with randomly generated absorbing states,  \emph{i.e.} random binary sequences where $1$'s and $0$'s occur with frequency $\rho$ and $1-\rho$, respectively; these have degeneracy $\binom{L}{\rho L}$. For the Manna model, $\rho_c < \rho_G$, hence the CID for random absorbing states is a smooth function around $\rho_c$. The inset shows how the absorbing states found by the dynamics have smaller CID and thus are more ordered than the uniformly sampled ones. Recent studies of the 2D Manna and related models \cite{PRL_hyper} indicate that they are hyperuniform \cite{torquato_local_2003} at the critical point, meaning that in this limit large-scale density fluctuations are anomalously suppressed.

\begin{figure*}
\includegraphics[width=\linewidth]{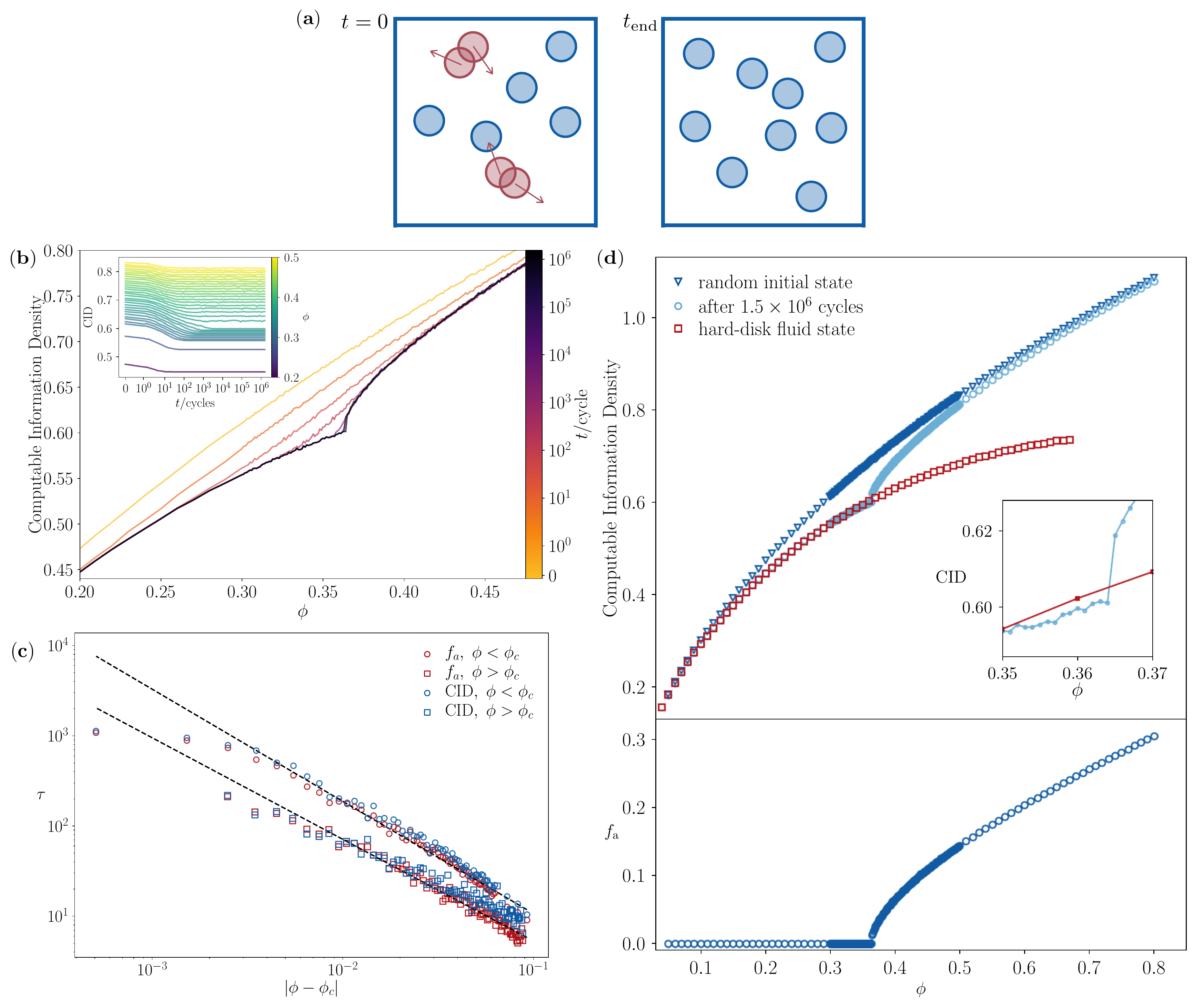}
\protect\caption{2D Random Organization model with $26,569 \times\phi$ disks of diameter $d$. Coordinates are quantized (digitized) using a square grid with bin size approximately $d/\sqrt{3}$. (a) At time $t=0$ the system is in a randomly sampled ideal gas configuration and disks are considered active when overlapping (active disks in red). At each time step a randomly selected active disk is given a random dispacement of size $\epsilon=d/3$, possible moves are illustrated by the red arrows. When the area fraction is below $\phi_c$, the system relaxes to an absorbing state, such that the fraction of active particles $f_a=0$. (b) CID as a function of area fraction $\phi$ for different times, a cycle corresponds to $N$ randomly attempted moves. The system orders as a function of time, developing a cusp at the critical area fraction $\phi_c \approx 0.364$. the inset shows the CID time evolution profile for several densities. (c) Characteristic time $\tau$ as a function of $|\phi - \phi_{c}|$, as measured by the decay of the fraction of active sites $f_{a}$ and the CID, showing  identical relaxation and critical exponent $\nu_\parallel = 1.2 \pm 0.2$ for both measures. Lines of best fit (dashed black lines) were obtained by bootstrapped minimum mean square error fits using a robust covariance estimator \cite{hubert2010minimum, efron1994introduction}. (d) Top panel: comparison of the random initial states, states found by the dynamics after $1.5 \times 10^6$ cycles and hard-disk fluid state corresponding to uniformly sampled absorbing states; the inset shows that near the critical point the absorbing states arrived at by the dynamics have lower CID, and are more ordered than the uniformly sampled ones. Lower panel: fraction of active sites as a function of $\phi$.\label{fig:4}}
\end{figure*}

The utility of the CID measure rests on the possibility to analyze experimental data, which do not, typically, lie on a lattice.  We therefore investigate a 2D continuum system, the ``random organization'' (RandOrg) model \cite{corte}, which was developed to explain the reversible-irreversible transition observed in experiments on sheared colloidal suspensions \cite{pine_chaos_2005}.  In RandOrg, the state of the system is given by positions of particles in real space, and in order to calculate the CID, the space must be discretized (quantized).  We choose a grid size such that there is at most one particle center in each box. Note that the resulting configurations is a coarse-grained representation of the original system and therefore the CID estimate may be subject to systematic deviations; we briefly discuss this issue in SI.

\begin{figure*}
\includegraphics[width=0.75\linewidth]{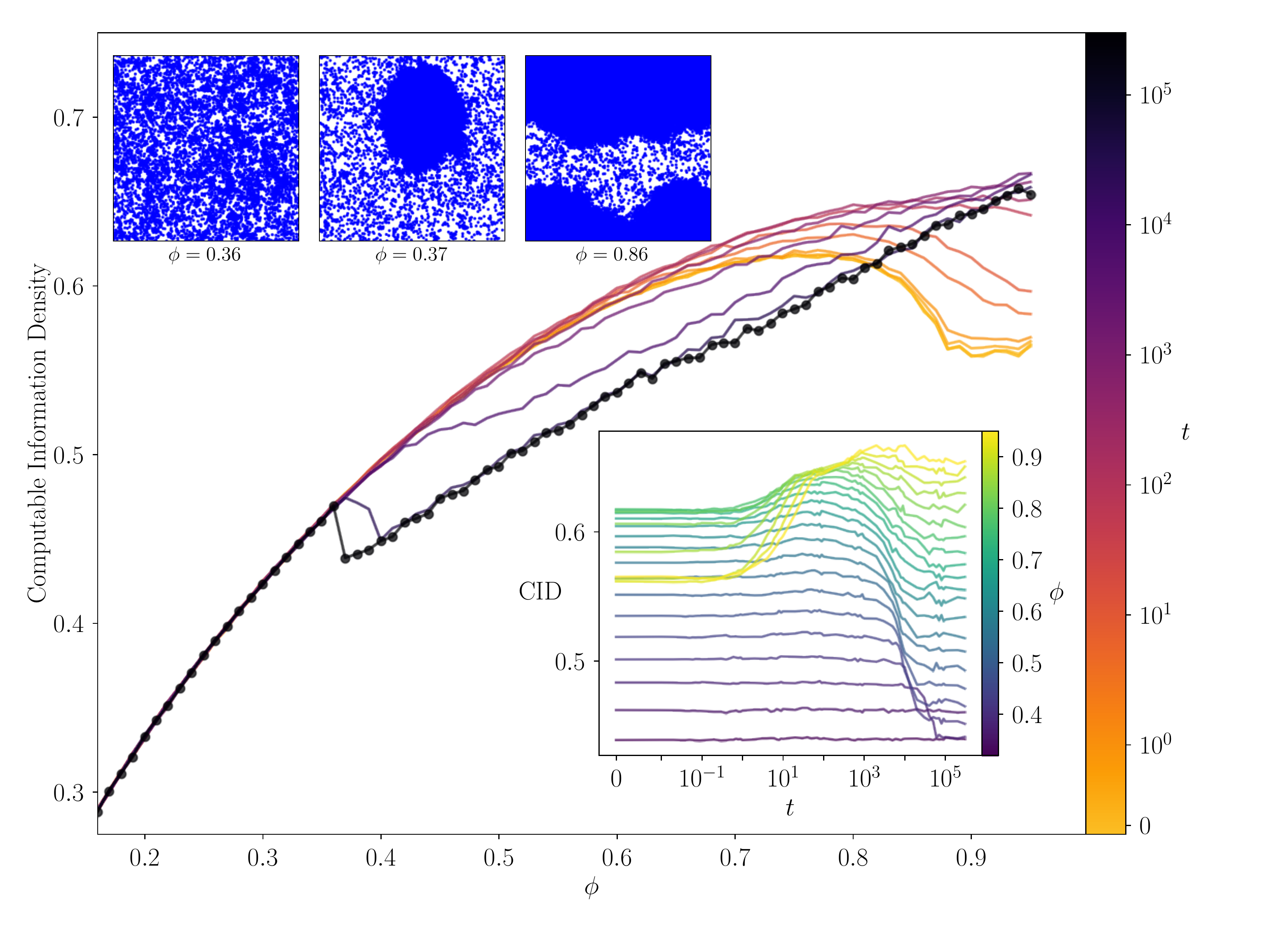}
\protect\caption{Active Brownian Particles:  Systems of 16,384 $\times\rho$  disks with short range repulsion, self-propelling at speed $v_0=0.1$.  Coordinates are quantized (digitized) using a square grid with bin-size approximately $d/\sqrt{5}$. At $\phi \approx 0.37$ the CID drops precipitously, indicating ordering associated with clustering and motility induced phase separation \cite{solon2018gt}. Representative configurations are shown for $\phi = 0.37, \, 0.39,\, 0.86$. For the initial quenched configurations (yellow curve) the flat region for $\phi \gtrapprox 0.88$ corresponds to samples consisting of small grain crystals. The inset shows the time dependence of the CID for different densities.
 \label{fig:5}}
\end{figure*}

In the simplest variant of RandOrg, identical disks are initially distributed randomly in space, with disks being considered active if they overlap.  At each time step a randomly selected active disk is given an independent random displacement $\vec{\epsilon}$, whose size $\epsilon$ is typically a fraction of a particle diameter. The control parameter for this model is the area fraction, $\phi = Na_0/A$, where $A$ is the area of the system and $a_0$ is the area of a particle. In this system $\phi_G = \phi_{\text{close packed}} \approx 0.91$, while {$\phi_c \approx 0.43$} in the limit $\epsilon \to 0$.   RandOrg can be considered a continuous version of the Manna model and has been shown to belong to the same universality classv\cite{menon2009universality}. An example of an initial random state and a final absorbing state is given in Fig.~\ref{fig:4}a.

We study RandOrg using a box of fixed area $A$, and generate initial configurations by randomly depositing $N=26569 \times \phi$ monodisperse disks with diameter $d$ for $256$ area fractions in the range $0.05 \leq \phi \leq 0.8$.  We then let the system evolve for approximately $1.5 \times 10^6$ full cycles with $\epsilon = d/3$ and periodic boundary conditions. For this system $\phi_c \approx 0.364$. We quantize the coordinates of the system using a square grid fine enough that the centers of two non-overlapping disks cannot occupy the same grid site and also require that the total number of bins be $2^{m} \times 2^m$, which is required by the Hilbert scan; in practice this results in a bin-size of approximately $d/\sqrt{3}$. 
 
In Fig.~\ref{fig:4}b we show the time evolution of the CID as a function of $\phi$ and in the inset we show the CID time evolution profiles for some of the area fractions; the curves are averaged over $7$ independent random initial conditions. As we would have expected by analogy with the Manna model, at long times the system develops a CID cusp around $\phi_c \approx 0.364$ and the critical slowing down is characterized in Fig.~\ref{fig:4}c. Note the remarkable agreement between the correlation times measured from the fraction of active particles $f_a$ and from the CID; we find $\nu_\parallel = 1.2 \pm 0.2$ on either side of the transition. In Fig~\ref{fig:4}d we compare the CID of RandOrg final stationary states with those of an equilibrium hard-disk fluid, corresponding to uniformly sampled absorbing states. The CID of the hard-disk fluid is smooth around $\phi_c$ because the hard disk fluid does not crystallize until the melting density $\phi_{\text{melting}} \approx 0.71$. The inset shows how the absorbing states found by the Manna dynamics are more ordered than those of the hard-disk fluid; just as for the Manna model, the absorbing states of RandOrg at the critical point are hyperuniform \cite{PRL_hyper}.

With our last example, we show that CID analysis is not limited to absorbing state models.  Here we consider a system of active Brownian particles exhibiting a motility-induced phase separation at a characteristic area fraction $\phi_c$; such behavior has been seen in experiments \cite{palacci2013living} and studied in theory \cite{fily2012athermal,fily2014freezing,solon2016generalized}.  The model by Fily and Marchetti \cite{fily2012athermal} consists of soft disks interacting via a short-ranged repulsive harmonic force $\mathbf{F}_{ij}=k(d-|\mathbf{r}_{ij}|)\Theta(d - |\mathbf{r}_{ij}|) \mathbf{r}_{ij}/|\mathbf{r}_{ij}|$ where $k$ is the spring constant, $d$ is the particle diameter, $\Theta$ is the Heavyside step function and $\mathbf{r}_{ij}=\mathbf{r}_{i} - \mathbf{r}_{j}$. The particles self-propel at fixed speed $v_0$ with orientation $\hat{\mathbf{n}}_i = \binom{\cos{\theta_i}}{\sin{\theta_i}}$.  The dynamics are overdamped, with mobility $\mu$ and zero-mean Gaussian rotational white noise $\eta_i(t)$ with rotational diffusion rate $\nu_r$, and are governed by the equations
\begin{equation}
\label{eq:eom}
\begin{aligned}
\partial_t \mathbf{r}_i &= v_0 \hat{\mathbf{n}}_i + \mu \sum_{j \neq i} \mathbf{F}_{ij} \\
\partial_t \theta_i &= \eta_i(t)
\end{aligned}
\end{equation} 

We prepare the system by depositing $N = 16384 \times \phi$ monodisperse disks in a fixed area for $95$ area fractions in the range $0.01 \leq \phi \leq 0.95$. We  minimize the energy by steepest descent \cite{wales2003energy}, and then let the system evolve under periodic boundary conditions with velocity $v_0=0.1$, mobility $\mu=1$, rotational diffusion rate $\nu_r=5\times10^{-4}$, and spring constant $k=1$.  We evolve the system according to Eq.~\ref{eq:eom} for time $t_{\mathrm{max}}=3 \times 10^5$ and time step $\Delta t = 10^{-2}$. We quantize the coordinates analogously to the protocol we followed for the RandOrg model, with a bin-size of approximately $d/\sqrt{5}$. In Fig.~\ref{fig:5} we show the CID as a function of area fraction at different times, and in the inset we show the CID time evolution profiles; curves are averaged over $6$ independent random initial configurations. 

At the lowest area fractions, the system is in an homogeneous gas-like state, and although the system is changing constantly with time, the CID remains unchanged from that of the initial non-overlapping random configurations. 
As seen in the inset, at $\phi_c \approx 0.37$ (the third curve from the bottom in the inset), the CID remains essentially constant until about $t \approx10^4$ iterations, when it drops, indicating the formation of a more ordered state.  Inspection of the configurations shown at the top of Fig.~\ref{fig:5} show this to be the result of a phase separation into dense liquid-like and less dense gas-like regions.  The step-like discontinuity in the CID between the initial time and the long-time steady state indicate a first order phase transition. These results confirm the density and velocity dependent phase transition previously reported \cite{fily2014freezing}, but present a clearer indication of the transition and clearly identify it as first order, in agreement with existing theoretical results \cite{solon2018generalized,solon2018gt}. In the SI we show results for a different velocity, which shifts the critical point.
At still higher densities, the CID is not monotonic in time, initially increasing before dropping. In this case the initial configurations, after relaxing by steepest decent to a configuration with no overlaps, are highly structured, and almost crystalline. When the particle activity is turned on, the order initially becomes disturbed.  At later stages the phase separation sets in and a different order sets in, reducing the CID. 

In this work we made a particular choice of universal code for data compression (LZ77), but other approaches are worth exploring, such as Kieffer-Yang grammar-based codes \cite{kieffer2000grammar, yang2000efficient, kieffer2000universal} and deep neural networks \cite{jiang1999image, hussain2018image}. Image compression techniques based on machine learning approaches have recently resurged due to improved methods for training deep networks \cite{gregor2016towards, toderici2017full, minnen2017spatially} and may inspire the development of better CID estimators.

The advent and use of powerful lossless data compression algorithms is a half century old. During this period, its application to many problems as well as its limitations have been extensively explored. Lossless compression not only provides a bound for entropy, but it is a surprisingly good one. The aim of this paper has been to illustrate that the CID provides a useful and readily implemented measure for systems out of equilibrium, accurately predicting critical points of phase transitions, their first or second order nature, and even yielding some critical exponents.  It allows a quantitative comparison of different states of a system and their time evolution, and it enables the discovery of new phases whose order can subsequently be characterized and studied.  These features give us reason to think that CID may find wide use in many areas of statistical many body physics, especially in the study of disordered and glassy systems, and make an important contribution to our understanding of correlation and organization.

\begin{acknowledgments}
We would like to thank Mark Adler, Ron Alfia, Daniel Hexner, Yariv Kafri, Johannes Klicpera, Yuval Lemberg, Neri Merhav and Emre Telatar for interesting and useful discussions. We are grateful to Ram Avinery, Roy Beck and Micha Kornreich for useful discussions and for providing their preprint on the application of data compression to study protein folding \cite{avinery2017universal}. This work was primarily supported by the National Science Foundation Physics of Living Systems Grant 1504867. DL thanks the US-Israel Binational Science Foundation (grant 2014713), the Israel Science Foundation (grant 1866/16), and the Initiative for the Theoretical Sciences at the Graduate Center of CUNY. P.M.C. was supported partially by the Materials Research Science and Engineering Center (MRSEC) Program of the National Science Foundation under Award DMR-1420073 
\end{acknowledgments}
\bibliography{bibliography}
\end{document}


\title{Supplementary Information: \\Quantifying hidden order out of equilibrium}
\author{Stefano Martiniani}
\email{smartiniani@nyu.edu}
\affiliation{Center for Soft Matter Research, Department of Physics, New York University, New York 10003, USA}
\author{Paul M. Chaikin} 
\email{chaikin@nyu.edu}
\affiliation{Center for Soft Matter Research, Department of Physics, New York University, New York 10003, USA}
\author{Dov Levine} 
\email{dovlevine19@gmail.com}
\affiliation{Department of Physics, Technion - IIT, 32000 Haifa, Israel}
\maketitle
\beginsupplement

\section{Lempel Ziv string-matching code (LZ77)}
Data sequences can be compressed by a large variety of universal codes; we do so by the `unrestricted' Lempel Ziv string-matching code, also known as LZ77 algorithm \cite{ziv1977universal, shields1999performance}.  Starting from the first character in the string, we decompose the sequence into `longest previous factors' (LPF), that are the longest subsequences that we encounter that have already occurred in the past. We represent each factor with the tuples $(i, \ell)$ where $i$ is the index/pointer to the position of the matching subsequence (or the character itself when it is observed for the first time) and $\ell$ is the length of the matching subsequence. 

The algorithm is best illustrated with an example. Consider the sequence $x=abcbabababab$ of length $L=12$ and alphabet $\alpha=\{a, b, c\}$ of size $|\alpha|=3$.  Taking the first character of the string to be at position 1, we get, for this example: $\text{LZ77}(x) = \{(a, 0), (b, 0), (c, 0), (2, 1), (1, 2), (5, 6)\}$. To understand this, we note that, at the outset, no factors have been identified, so that the first factor is the first character, $x_1=a$ and the length of the matching subsequence is obviously $0$, hence the LPF is $(a, 0)$.  Moving one position to the right, we encounter $x_2=b$, which has not yet been seen, and likewise for the next character, $x_3=c$, thus we have LPF $(b, 0)$ and $(c, 0)$, respectively.  Moving to the next position $4$, we see that of the subsequences starting at this position, ($b$, $ba$, $bab$, ...), only the single character subsequence $b$ has been encountered (at position 2); and the LPF is $(2,1)$ instructing the decoder to copy 1 character starting at position 2.  Starting at the following position $5$, we note that the words $a$ and $ab$ have already occurred (starting at position 1), but not $aba, abab, \dots$ thus the LPF is $(1,2)$ instructing the decoder to copy 2 characters beginning at position 1.  Moving to position 7 we note that the entire remaining string, $ababab$, corresponds to the previous subsequence $x_5x_6=ab$ copied cyclically for 6 characters, thus the LPF is $(5,6)$.  This gives us a list of $C = 6$ tuples with which  the entire original string may be reconstructed; it is the LZ77 encoding of the sequence.  

The total binary code length $\mathcal{L}(x)$ of the LZ77 encoding can be computed from the number $C$ of longest previous factors: It takes $\log(|\alpha| + L)$ bits to specify a position in the sequence $x$ or a location in the dictionary $\alpha$, and for a prefix code it takes $\log \ell_j + O(\log \log \ell_j)$ bits to specify $\ell_j$, the length of the matching subsequence for the $j$-th factor \cite{shields1999performance}. Hence the total binary code length is bounded as
\begin{equation}
\begin{aligned}
\mathcal{L}_{\text{LZ77}}(x) &\leq C \log (L + |\alpha|) + \sum_{j=1}^{C} \log \ell_j + O\left( \sum_{j=1}^C \log \log \ell_j \right) \\
&\leq C \log C+ 2C \log \frac{L}{C} + O \left( C \log \log \frac{L}{C} \right),
\label{eq:lzc}
\end{aligned}
\end{equation}
where the final bound was obtained by concavity of the log (Jensen's inequality), we assumed $L \gg |\alpha|$, and all $\log$ are base $2$ throughout. The CID is simply the ratio
\begin{equation}
\text{CID} \equiv \frac{\mathcal{L}(x)}{L}.
\end{equation}
Note that the CID is not the same as the compression factor (or compressibility) \cite{ziv1978compression}
\begin{equation}
\varrho = \frac{\text{CID}}{\log |\alpha|},
\label{eq:compression_factor}
\end{equation}
corresponding to the amount of information per character of the \textit{binary} representation of the uncompressed sequence $x$, although they are equivalent for binary sequences (when $|\alpha|=2$). Thus, we have that $0 \leq \varrho \leq 1$ while the $\text{CID} \geq 0$ is not bounded from above and it is indeed an information `density'.

\section{Rate of convergence}

How well a code compresses a sequence is measured in terms of the `\textit{redundancy}'
\begin{equation}
R \equiv \mathop{\mathbb{E}}\left(\frac{\mathcal{L}(x)}{L} \right) - H,
\end{equation}
that is the amount by which the average $\mathrm{CID} \equiv \mathcal{L}(x)/L $ exceeds the entropy (per character) of the source. Shannon had demonstrated that the redundancy cannot be negative and there exist an \textit{optimal} code for which the redundancy is zero \cite{shannon2001mathematical}. It can be shown that when the sequence $x$ is sampled from a stationary and ergodic process, LZ codes achieve optimal compression, hence $R \to 0$ as $L \to \infty$, and for individual deterministic sequences LZ77 codes do at least as well as the empirical (block) entropy, if not better \cite{shields1999performance}. 

The rate at which optimality can be attained is rather slow, in general $\sim 1/\log L$. Knowledge of the precise redundancy rate would allow an effortless extrapolation to the thermodynamic limit ($L \to \infty$); unfortunately, such precise bounds are not known in general \cite{shields1993universal} and the question of how to best extrapolate finite-size measurements of the LZ complexity is still open. In what follows we analyse two examples for which the exact value of the entropy is known analytically, namely a Bernoulli sequence and the two-dimensional Ising model, and try to gain some insight into the rate of convergence of LZ77, on the basis of numerical results and some already established theoretical results.

\subsection{Redundancy bounds of Markov sources}

The general class of unifilar Markov sources includes the models for which each output depends statistically on the last $l$ symbols. Savari \cite{savari1998redundancy} obtained the following bounds for positive entropy ($H>0$) and zero entropy ($H=0$) unifilar Markov sources
\begin{align}
R_{\text{LZ77}}^{(H>0)} &\leq A H \frac{\log\log L}{\log L} + o\left(\frac{\log\log L}{\log L}\right)  \label{eq:bound_positive}\\
R_{\text{LZ77}}^{(H=0)} &\leq B \frac{\log L}{L} + O\left( \frac{\log\log L}{\log L}\right),
\label{eq:bound_zero}
\end{align}
with $A=2$ and $B = 2(S+1)$, where $S$ is the number of states of the Markov source \footnote{Note that the big-O notation $g(n)=O(f(n))$ means that $g$ and $f$ are of the same order, hence $\lim_{n \to \infty} g(n)/f(n) = c$ where $c$ is a constant; while the little-o notation $g(n)=o(f(n))$ means that $g$ is ultimately smaller than $f$, hence $\lim_{n \to \infty} g(n)/f(n) = 0$. }.

\subsection{Deterministic sequences}

\begin{figure}	
\centering
\includegraphics[width=\linewidth]{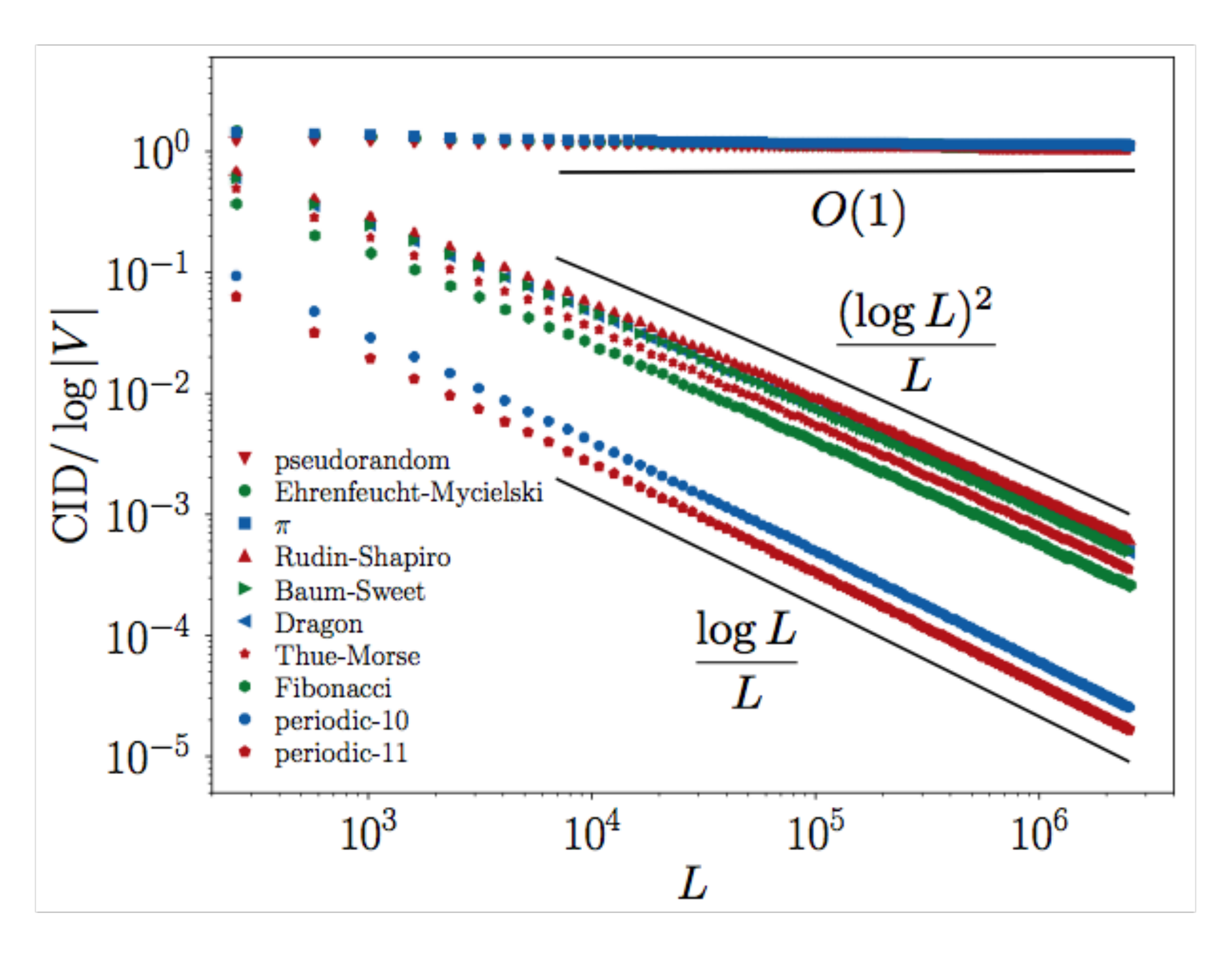}
\caption{\label{fig:det_sequences}  CID of deterministic sequences of length $L$ computed by LZ77 and normalized by $\log |\alpha|$, where $|\alpha|$ is the alphabet size (defined as compression factor in Eq.~\ref{eq:compression_factor}). OEIS \cite{oeis2018} sequence numbers are reported along with the sequence names. We consider three types of sequences. Periodic sequences ($\sim \log L / L$): a periodic sequence with unit `10' ( \textit{periodic-10}) and a sequence of all `1' (\textit{periodic-11}). Quasiperiodic sequences (fixed points of morphisms $\sim (\log L)^2 / L$): \textit{Rudin-Shapiro} (A020985); \textit{Baum-Sweet} (A037011); \textit{Dragon} (A014577); Thue-Morse (A010060); \textit{Fibonacci word} (A003849). Pseudorandom sequences ($\sim O(1)$): a \textit{pseudorandom} binary sequence generated using the Mersenne Twister pseudorandom number generator \cite{matsumoto1998mersenne}; \textit{Ehrenfeucht-Mycielski} (A007061); and the digits of $\pi$ (A000796).} 
\end{figure}

In Fig.~\ref{fig:det_sequences} we show the CID computed by LZ77 for a number of deterministic sequences. We can divide the sequences into three groups: (i) the periodic sequences that converge to zero entropy as $\sim \log L / L$, in agreement with Savari \cite{savari1998redundancy}; (ii) the quasiperiodic sequences (fixed points of morphisms) that converge to zero as $\sim (\log L)^2 / L$, in agreement with Constatinescu and Ilie \cite{constantinescu2007lempel}; (iii) deterministic but statistically pseudorandom sequences that approach a plateau for large $L$, these include the digits of $\pi$ (sequence A000796 of OEIS \cite{oeis2018}) and the `maximally unpredictable' Ehrenfeucht-Mycielski sequence (sequence A007061 of OEIS \cite{oeis2018}). It is clear that the CID as measured by LZ77 despite being an algorithmic complexity is rather similar to the empirical (block) entropy in that its value depends on the statistical properties of the sequence and usually cannot discern finite pseudorandom sequences from truly random ones, despite they can be constructed by a relatively simple algorithm and therefore have Kolmogorov complexity per character $K=0$.

\subsection{Extrapolation for $H>0$ source}

As we have seen LZ77 converges quickly ($\sim \log L /L$) for both periodic and quasiperiodic sequences. However, when the source has positive entropy the rate of convergence becomes exponentially slow ($\sim \log \log L / \log L$), meaning that a careful extrapolation to the thermodynamic limit $L \to \infty$ is necessary. 

Let us denote $\hat{H}_L \equiv \mathbb{E}(\text{CID(x)})$, then by rearranging Eq.~\ref{eq:bound_positive}  we can write an estimator for $H$
\begin{equation}
H = \frac{\hat{H}_L  \log L}{\log L + A^{\star} \log \log L} \text{ as } L \to \infty,
\label{eq:raw_estimator}
\end{equation}
where we have replaced the inequality with an equality by substituting $A$ (the asymptotic upper bound) with an effective $A^\star$, which is simply the rate at which the relative error goes to zero, since
\begin{equation}
\frac{\hat{H}_L - H}{H} = A^\star \frac{\log\log L}{\log L} \text{ as } L \to \infty.
\label{eq:rel_redundancy}
\end{equation}

The obvious approach to obtain $H$ is a direct extrapolation to $\log \log L / \log L \to 0$ by finite size scaling analysis, so that 
\begin{equation}
\hat{H}_L = \hat{H}_\infty + \hat{A}^\star \hat{H}_\infty \frac{\log \log L}{\log L},
\label{eq:fss_extrapolation}
\end{equation}
where the intercept $\hat{H}_\infty$ is our estimate for $H$ and $\hat{A}^\star$ is our estimate for $A^\star$, that is unique to the source. Alternatively, if the asymptotic value of the entropy is known, $\hat{A}^\star$ can be estimated directly from Eq.~\ref{eq:raw_estimator} for finite $L$. Note that if we were to compute the entropy of a thermal equilibrium system then we would need to fit $\hat{H}_\infty$ and $\hat{A}^\star$ for each temperature.

\begin{figure*}
\begin{subfigure}{0.5\textwidth}
  \centering
  \topinset{\bf(a)}{\includegraphics[width=\linewidth]{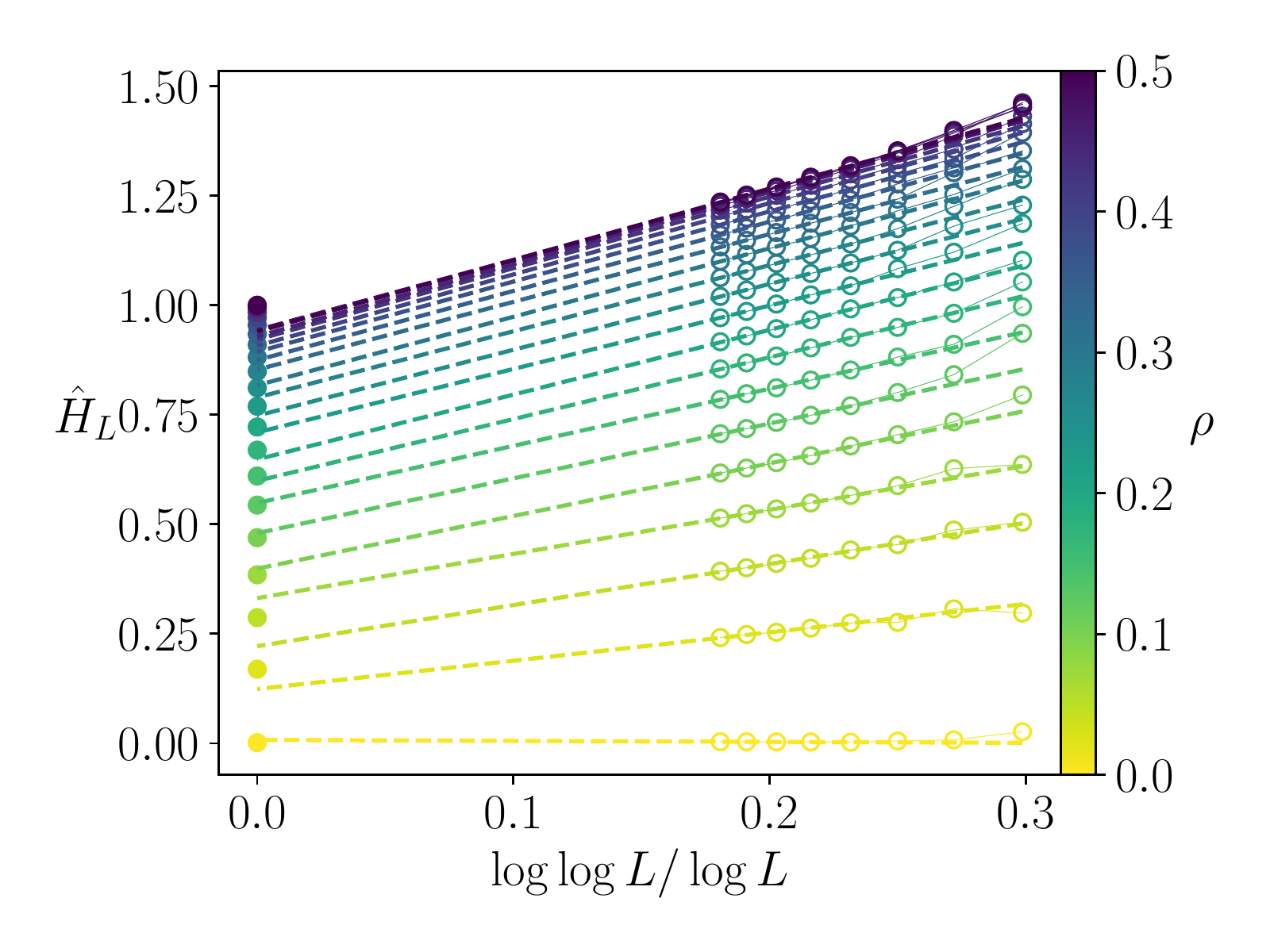}}{0.15in}{.08in}
\end{subfigure}%
\begin{subfigure}{0.5\textwidth}
  \centering
   \topinset{\bf(b)}{\includegraphics[width=\linewidth]{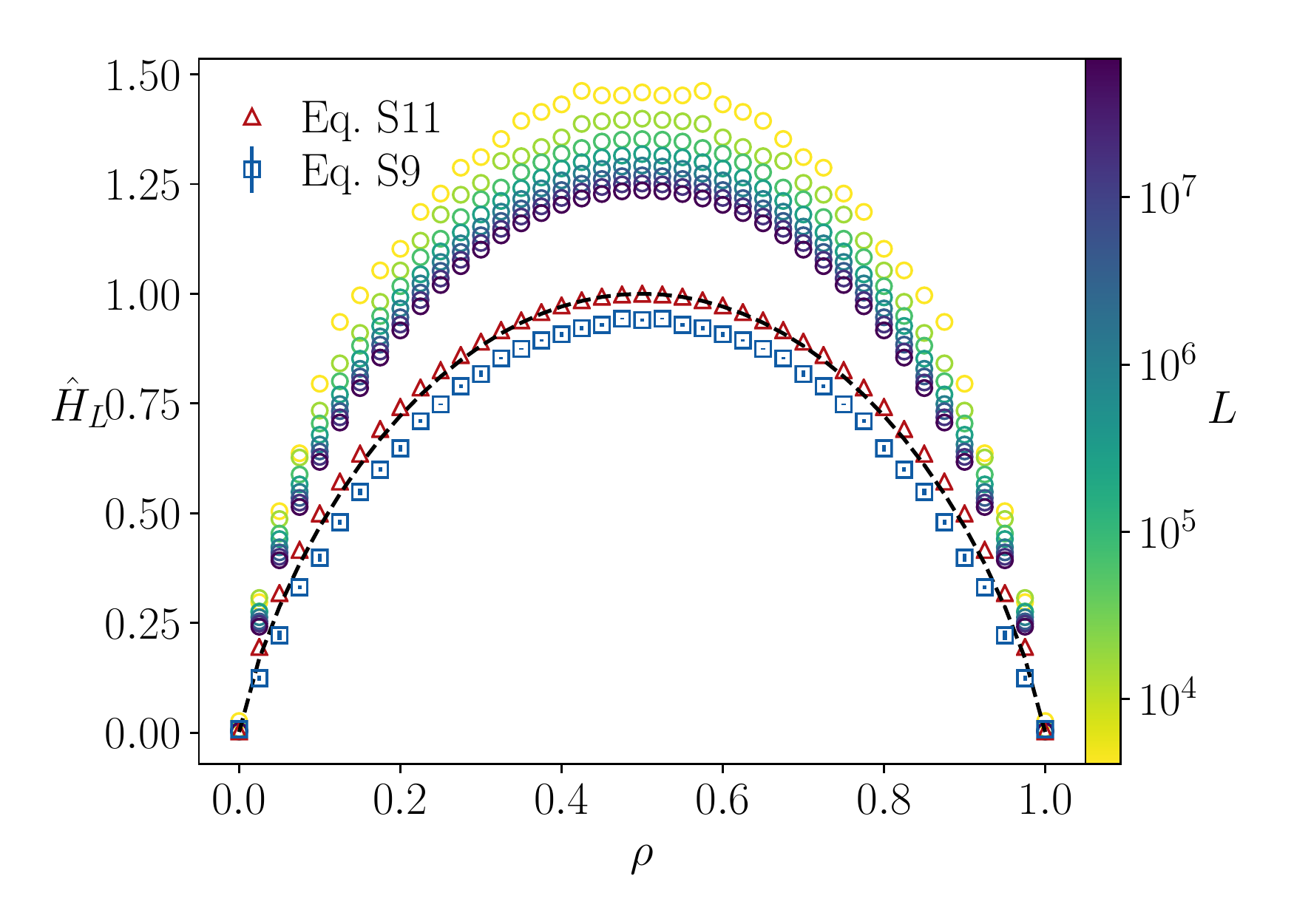}}{0.15in}{.08in}
\end{subfigure}
\begin{subfigure}{0.5\textwidth}
  \centering
    \topinset{\bf(c)}{\includegraphics[width=\linewidth]{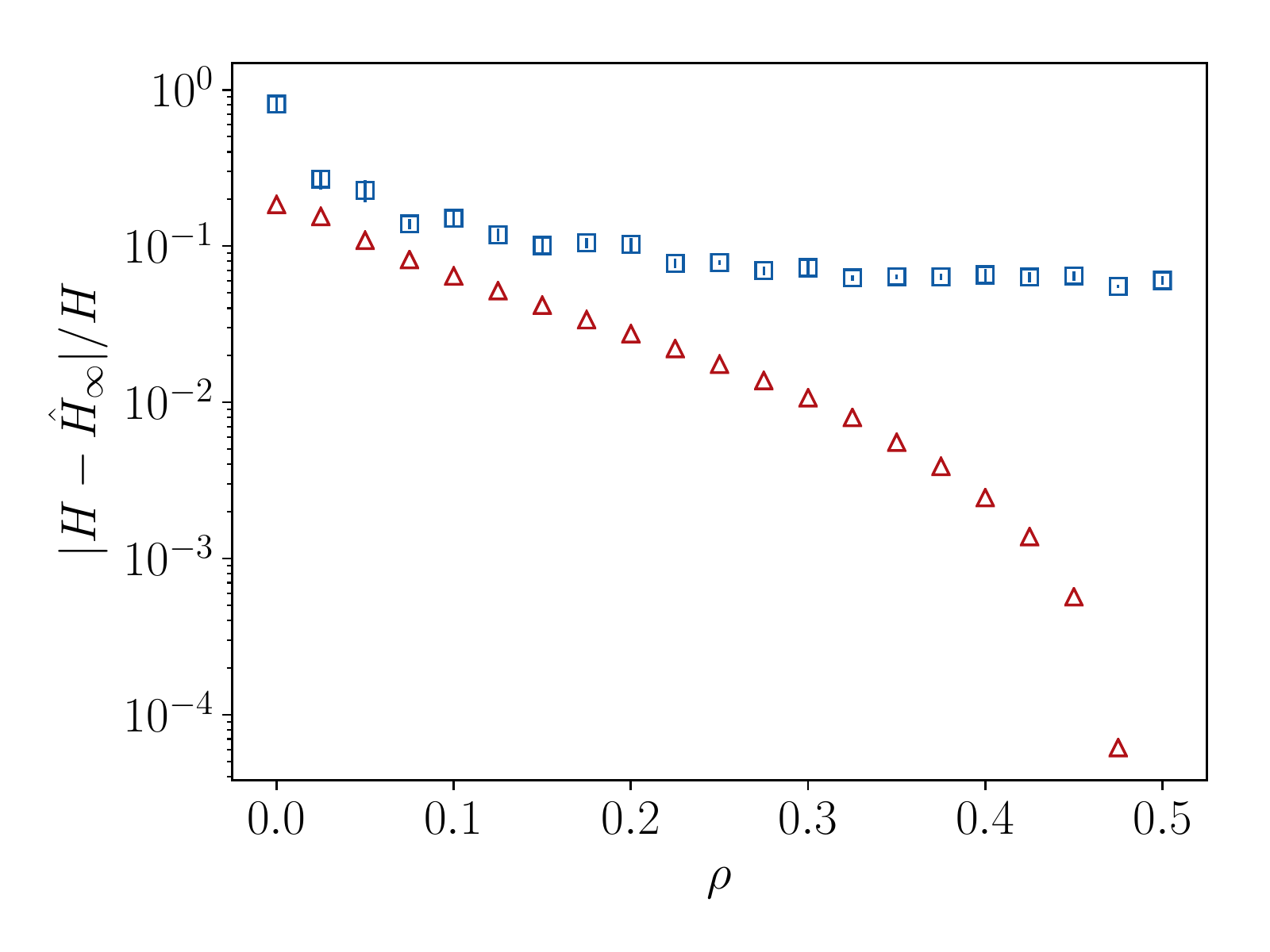}}{0.15in}{.08in}
\end{subfigure}%
\begin{subfigure}{0.5\textwidth}
  \centering
    \topinset{\bf(d)}{\includegraphics[width=\linewidth]{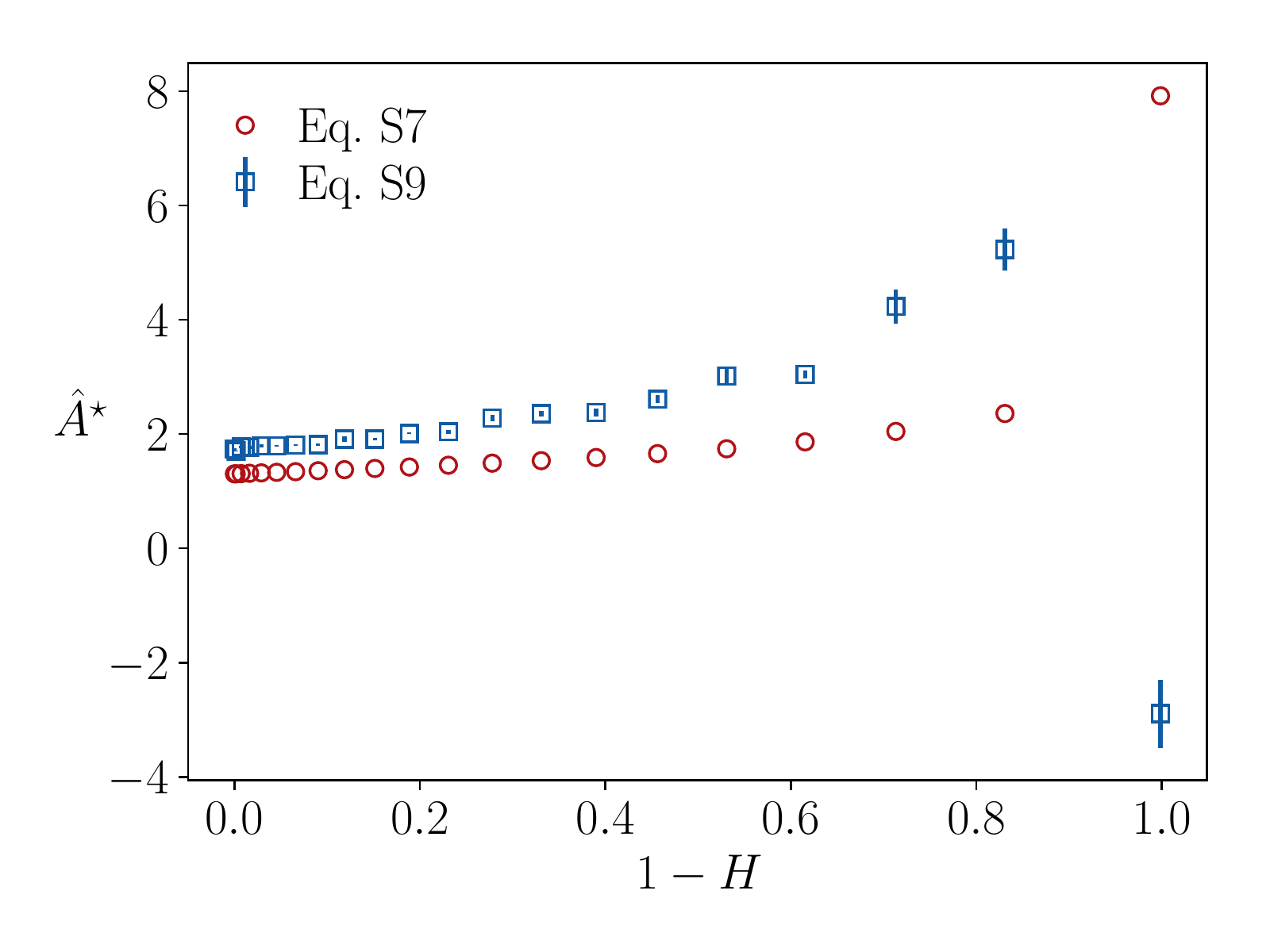}}{0.15in}{.08in}
\end{subfigure}
\caption{\label{fig:bernoulli} Analysis of Bernoulli sequences generated for $21$ values of $\rho$ in the range $0 < \rho \leq 0.5$ for $8$ different system sizes $L=2^{2m}$ with $6 \leq m \leq 13$, and computed the CID by LZ77. (a) Empty circles correspond to CID measurements with the corresponding  $\rho$ indicated by the colorbar. The filled circles correspond to the exact value in the limit $L \to \infty$. Solid lines connecting the circles are a guide to the eye while the dashed lines correspond to linear fits of the 5 largest sizes for each $\rho$. (b) Empty circles correspond to CID measurements with the corresponding system size $L$ indicated by the colorbar. The dashed black line is the exact value in the limit $L \to \infty$. The red triangles correspond to $\hat{H}_\infty$ as computed by Eq.~\ref{eq:norm_extrapolation} and the blue squares as computed by Eq.~\ref{eq:fss_extrapolation}. (c) Relative error as a function of $\rho$ for  $\hat{H}_\infty$ as computed by Eq.~\ref{eq:norm_extrapolation} (red triangles) and Eq.~\ref{eq:fss_extrapolation} (blue squares). (d) Effective relative rate of convergence $\hat{A}^\star$ estimated from Eq.~\ref{eq:raw_estimator} (red circles) using the exact asymptotic value of the entropy and from Eq.~\ref{eq:fss_extrapolation} (blue squares) as in (a).}
\end{figure*}

\begin{figure*}
\begin{subfigure}{0.5\textwidth}
  \centering
  \topinset{\bf(a)}{\includegraphics[width=\linewidth]{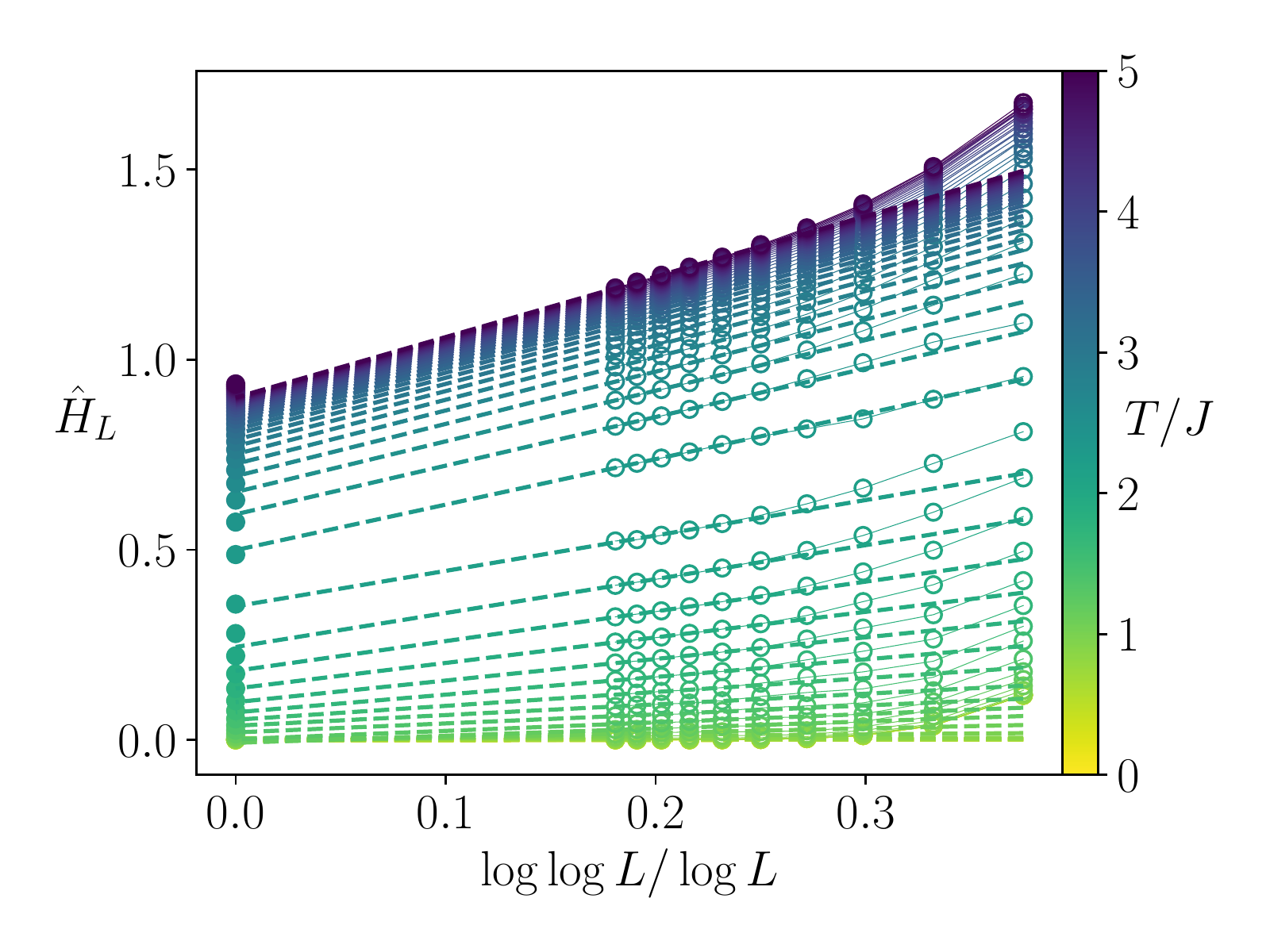}}{0.15in}{.08in}
\end{subfigure}%
\begin{subfigure}{0.5\textwidth}
  \centering
  \topinset{\bf(b)}{\includegraphics[width=\linewidth]{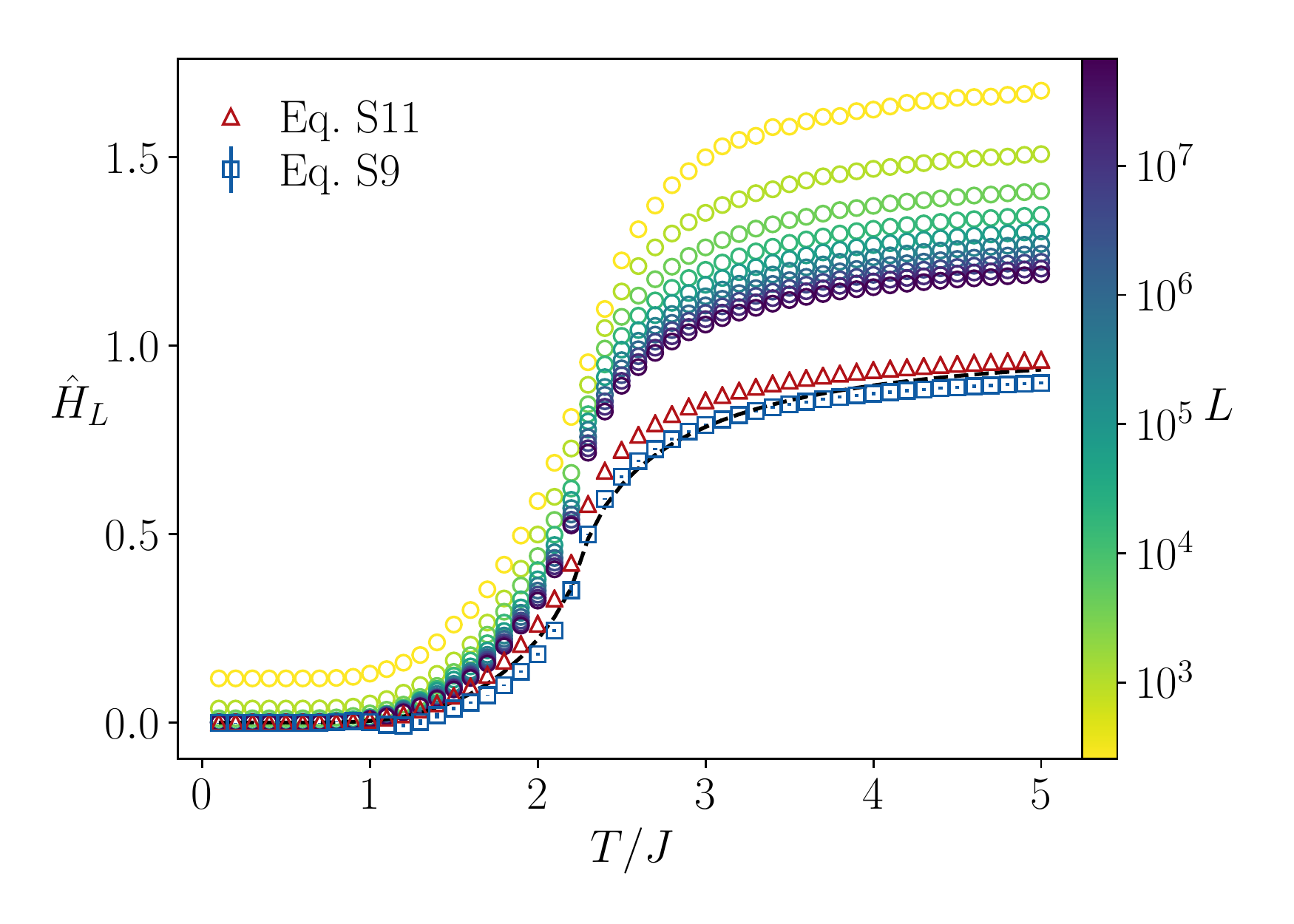}}{0.15in}{.08in}
\end{subfigure}
\begin{subfigure}{0.5\textwidth}
  \centering
    \topinset{\bf(c)}{\includegraphics[width=\linewidth]{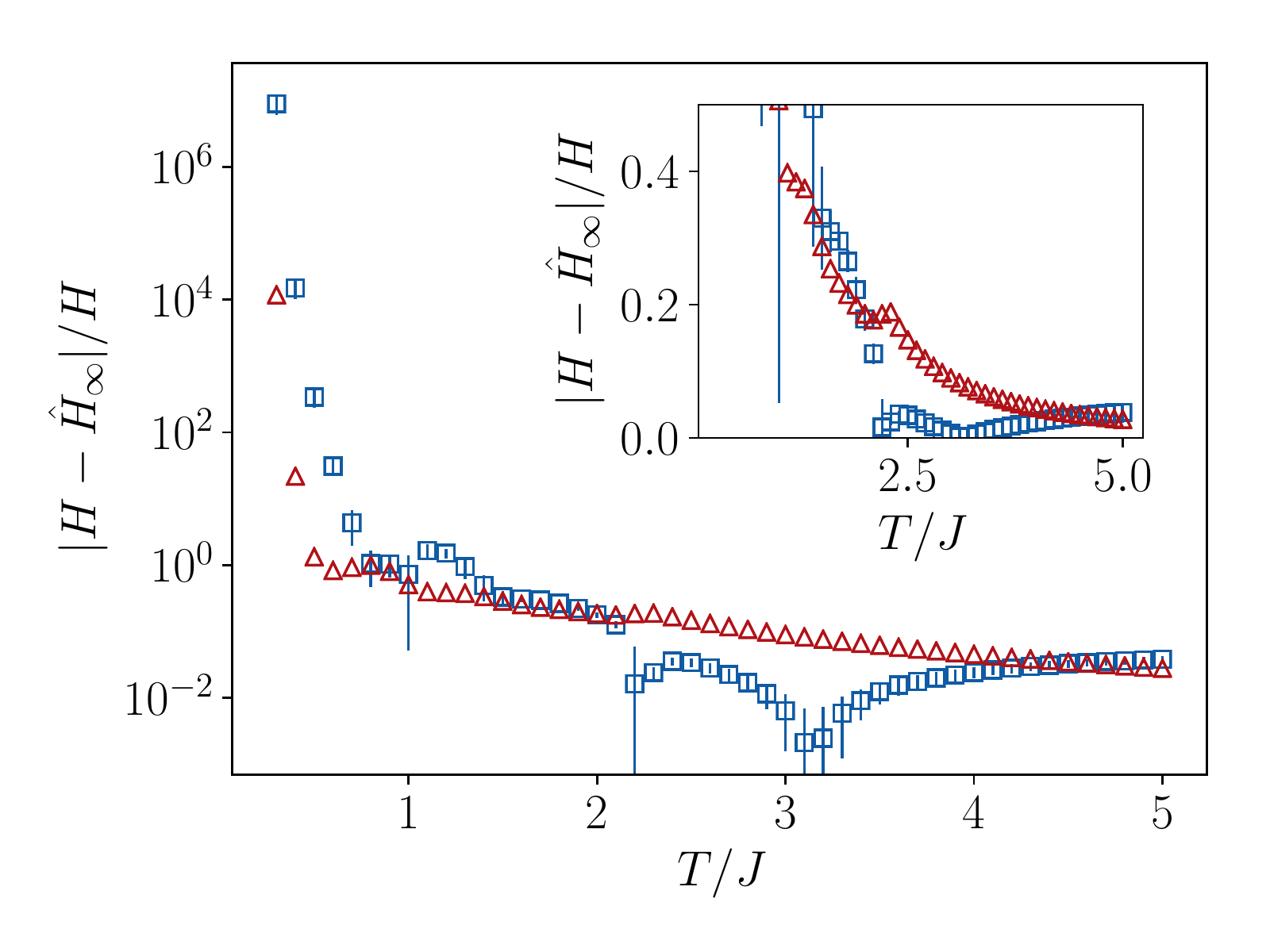}}{0.15in}{.08in}
\end{subfigure}%
\begin{subfigure}{0.5\textwidth}
  \centering
    \topinset{\bf(d)}{\includegraphics[width=\linewidth]{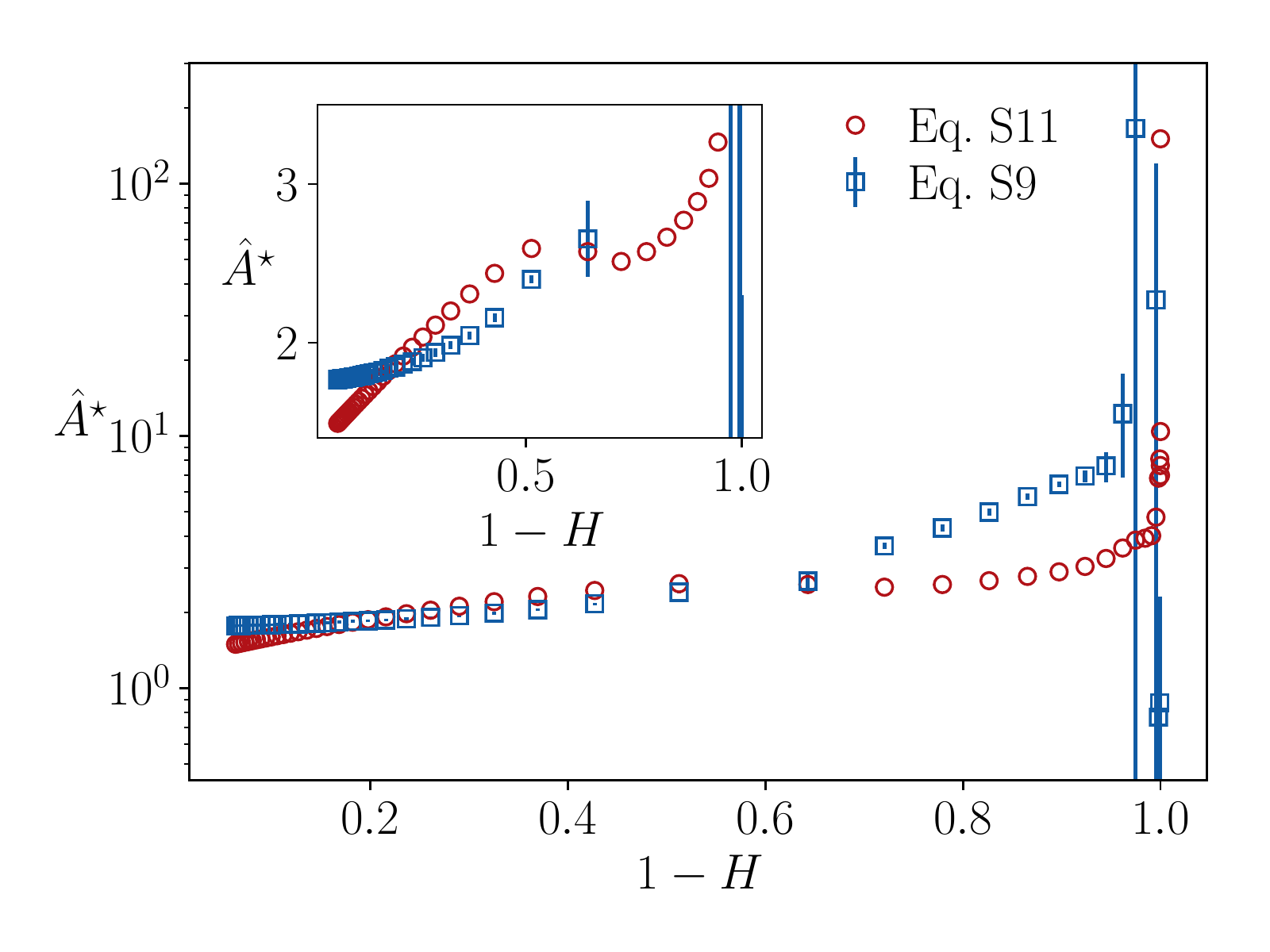}}{0.15in}{.08in}
\end{subfigure}
\caption{\label{fig:ising} Analysis of two dimensional ferromagnetic Ising model with no external field simulated by Wolff dynamics \cite{wolff1989collective}. We simulated the two-dimensional Ising model for 50 temperatures in the range $0.1 \leq T/J \leq 5$ for $10$ different system sizes $L=2^m \times 2^m$ with $4 \leq m \leq 13$, scanned the configurations according to a Hilbert curve and computed the CID by LZ77. (a) Empty circles correspond to CID measurements with the corresponding temperature $T/J$ indicated by the colorbar. The filled circles correspond to the exact value in the limit $L \to \infty$. Solid lines connecting the circles are a guide to the eye while the dashed lines correspond to linear fits of the 5 largest sizes for each temperature. (b) Empty circles correspond to CID measurements with the corresponding system size $L$ indicated by the colorbar. The dashed black line is the exact value in the limit $L \to \infty$. The red triangles correspond to $\hat{H}_\infty$ as computed by Eq.~\ref{eq:norm_extrapolation} and the blue squares as computed by Eq.~\ref{eq:fss_extrapolation}. (c) Relative error as a function of $T/J$ for  $\hat{H}_\infty$ as computed by Eq.~\ref{eq:norm_extrapolation} (red triangles) and Eq.~\ref{eq:fss_extrapolation} (blue squares). Inset shows details of the curves in a linear plot. (d) Effective relative rate of convergence $\hat{A}^\star$ estimated from Eq.~\ref{eq:raw_estimator} (red circles) using the exact asymptotic value of the entropy and from Eq.~\ref{eq:fss_extrapolation} (blue squares) as in (a). Inset shows details of the curves in a linear plot.}	
\end{figure*}

The second approach that we propose, and the method of choice in the main text, is to use a lower bound for $\hat{A}^\star$ computed from random binary sequences. The simplest way of doing so is to use the fact that $\hat{A}^\star$ should be independent of $L$ when this is large, then from Eq.~\ref{eq:raw_estimator} we can write
\begin{equation}
 \hat{A}^{{\star}^{\text{(rand)}}} = (\hat{H}_L^{\text{(rand)}} - 1) \frac{\log L}{\log \log L},
\end{equation}
where $\hat{H}_L^{\text{(rand)}}$ is the average CID for a random binary sequence and we have used the fact that $H^{\text{(rand)}} = 1$. This choice is equivalent to the following bound estimator
\begin{equation}
\hat{H}_\infty \leq \frac{\hat{H}_L}{\hat{H}_L^{\text{(rand)}}}.
\label{eq:norm_extrapolation}
\end{equation}
In Figs.~\ref{fig:bernoulli}d,\ref{fig:ising}d we verify that $\hat{A}^\star$ is minimal for binary random sequences.

\subsection{Numerical results}
We perform finite size scaling analysis for two model systems: (i) Bernoulli random sequences, defined so that each element of the sequence is i.i.d. with value $1$ with probability $\rho$ and $0$ with probability $1-\rho$, (ii) the two-dimensional Ising model with nearest-neighbour ferromagnetic interactions ($J>0$) and no external field, simulated by Wolff dynamics \cite{wolff1989collective} at temperature $T$. 

\subsubsection{Bernoulli sequences}

We generated Bernoulli sequences for $21$ values of $\rho$ in the range $0 < \rho \leq 0.5$ for $8$ different system sizes $L=2^{2m}$ with $6 \leq m \leq 13$, and computed the CID by LZ77. In Fig.~\ref{fig:bernoulli}a we show the extrapolation according to Eq.~\ref{eq:fss_extrapolation}. In Fig.~\ref{fig:bernoulli}b we show the estimated entropy $\hat{H}_L$ as a function of system size against the theoretical expectation (dashed line) and the extrapolated values $\hat{H}_\infty$ by Eq.~\ref{eq:fss_extrapolation} (blue squares) and Eq.~\ref{eq:norm_extrapolation} (red triangles) for the largest system size $L=2^{26}$. Normalising by the entropy of a random binary sequence (Eq.~\ref{eq:norm_extrapolation}) clearly gives more accurate results than a direct extrapolation (Eq.~\ref{eq:fss_extrapolation}), as it can be inferred by the relative error in Fig.~\ref{fig:bernoulli}c. In Fig.~\ref{fig:bernoulli}d we show that $\hat{A}^\star$ is minimal for high entropy and monotonically increasing with decreasing entropy. Also note that $\hat{A}^\star \leq 2$ over almost the whole range, as suggested by Savari, until $H \to 0$ and $\hat{A}^\star$ diverges (note that in this limit the extrapolation should be done according to Eq.~\ref{eq:bound_zero}).

\subsubsection{Ising model}

We simulated the two-dimensional Ising model for 50 temperatures in the range $0.1 \leq T/J \leq 5$ for $10$ different system sizes $L=2^m \times 2^m$ with $4 \leq m \leq 13$, scanned the configurations according to a Hilbert curve and computed the CID by LZ77. In Fig.~\ref{fig:ising}a we show the extrapolation according to Eq.~\ref{eq:fss_extrapolation}. In Fig.~\ref{fig:ising}b we show the estimated entropy $\hat{H}_L$ as a function of system size against the theoretical expectation (dashed line) and the extrapolated values $\hat{H}_\infty$ by Eq.~\ref{eq:fss_extrapolation} (blue squares) and Eq.~\ref{eq:norm_extrapolation} (red triangles) for the largest system size $L=2^{26}$. In this case normalising by the entropy of a binary random sequence (Eq.~\ref{eq:norm_extrapolation}) does not necessarily yield a lower error, though it is preferrable because more stable across temperatures, as it can be seen in Fig.~\ref{fig:ising}c.  In Fig.~\ref{fig:ising}d we show that $\hat{A}^\star$ is minimal for high entropy and roughly monotonically increasing with decreasing entropy, with a small deviation from monotonicity near the critical point (the details of this feature depend on the system size, so it is not clear what its origin is). In this case the asymptotic upper bound $\hat{A}^\star \leq 2$ does not quite hold, although this is probably a finite size effect since as $L$ grows, $\hat{A}^\star$ consistently decreases. Note that for this system we expect finite size effects to be enhanced by the scanning of a two-dimensional structure into a one-dimensional sequence.

\begin{figure}	
\centering
\includegraphics[width=\linewidth]{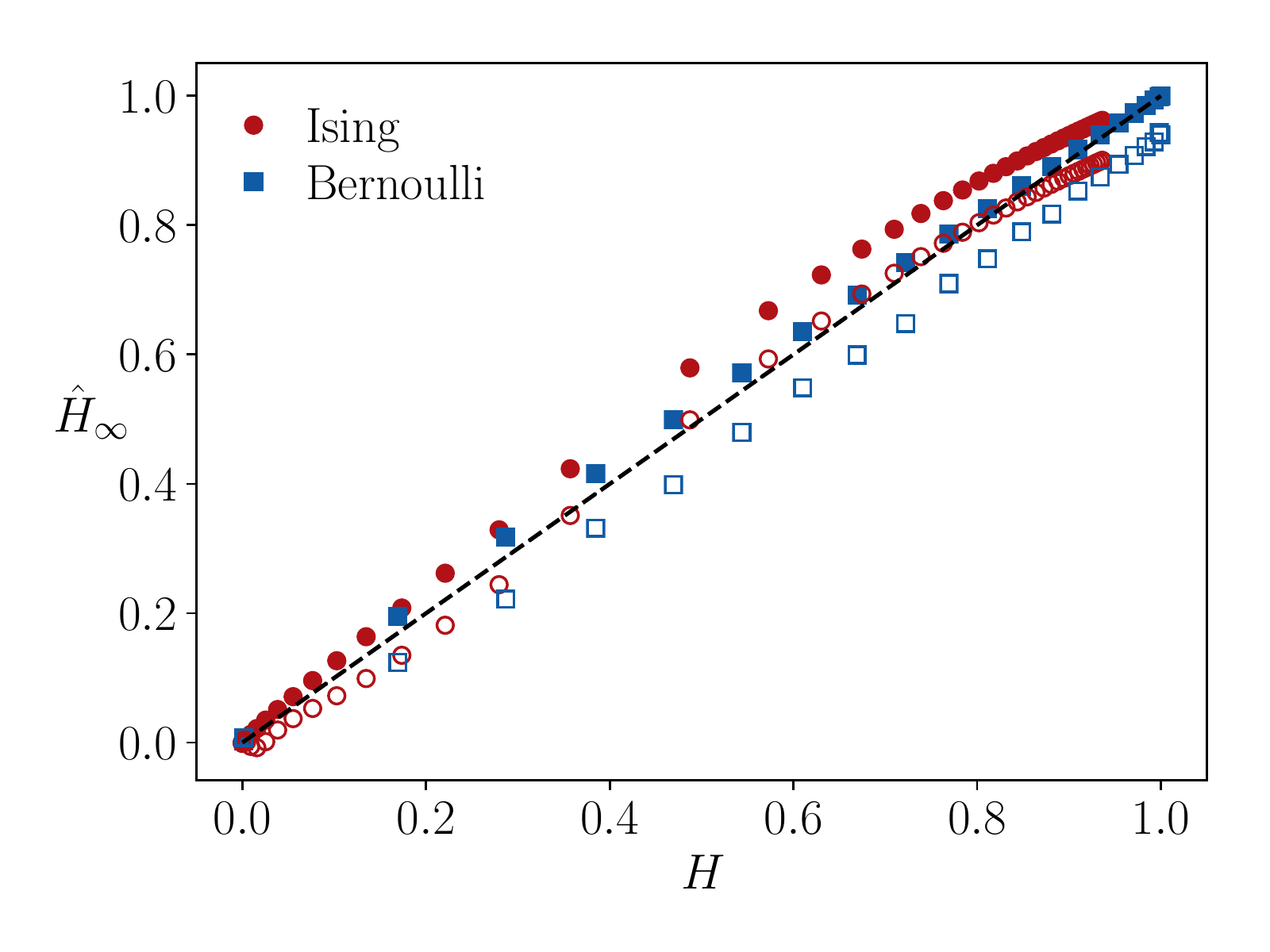}
\caption{\label{fig:bern_vs_ising}  Comparison of estimated entropies $\hat{H}_\infty$ as computed by Eq.~\ref{eq:norm_extrapolation} (filled markers) and Eq.~\ref{eq:fss_extrapolation} (void markers), as a function of the theoretical expected entropy $H$. Red circles correspond to measurements for the Ising model and blue squares to measurements for the Bernoulli sequences, both for $L=2^{26}$. The black dashed line is the curve $\hat{H}_\infty=H$.}
\end{figure}

Finally, in Fig.~\ref{fig:bern_vs_ising} we compare the entropies $\hat{H}_\infty$ estimated by the two proposed approaches, Eq.~\ref{eq:fss_extrapolation} and Eq.~\ref{eq:norm_extrapolation}, against the theoretical expectation $H$. Though the estimates are all in the vicinity of the expected value, as well as monotonic for these systems, for a given $H$ the extrapolated values $\hat{H}_\infty$ for a given method do not always agree between the two systems, since sources with the same entropy may converge at different rates.

\section{Continuum systems}

One of the more exciting applications of the CID is to continuum systems. For this class of problems we must discretise the system's degrees of freedom (e.g. the particle coordinates) according to some protocol, for instance according to a square or an hexagonal grid, before they can be compressed. 

Consider a quantization of the support of a random variable $X$ in $n$ bins, and let us denote the quantized (discrete) random variable as $X^{(n)}$. The analogue of the entropy for a continuous distribution is known as \textit{differential entropy} 
\begin{equation}
h(X) = -\int p(x) \log p(x) \mathrm{d}x
\end{equation}
and to determine $X$ with $\log n$ bits of accuracy we require on average $H(X^{(n)}) \approx h(X) + \log n$ bits of information \cite{cover2012elements}. This means that specifying a continuous random variable to arbitrary precision is not possible as it requires an infinite amount of information. 

In addition, the resulting configuration is a coarse-grained representation of the original and the CID estimation may be subject to systematic deviations, due for instance to the convolution of a square grid with the system's coordinates. These effects are known in information theory as ``rate distortion'' \cite{cover2012elements}. Put simply, despite using a lossless data compression algorithm, through quantization of a continuous configuration, we inevitably lose some information and introduce systematic errors. Physically, this is a coarse-graining issue and the choice of the the best protocol to adopt depends on the problem at hand. The possibility of developing a protocol that minimizes rate-distortion in an unsupervised fashion for the kind of problems we consider remains to be explored.

\section{Implementation details}

Numerical simulations were performed using the open source libraries \textit{Pele} \cite{pele} and \textit{MCPele} \cite{mcpele}. LZ77 compression was performed using the open source library \textit{Sweetsourcod} \cite{sweetsourcod} wrapping the linear time algorithms for LZ77 by Karkkainen, Kempa and Puglisi \cite{kempa2013lempel, karkkainen2013linear, karkkainen2016lazy}. We adopt the KKP2 algorithm from \cite{kkp}, capable of performing the LZ77 factorization in $O(N)$ time complexity. The implementation of the Hilbert curve in \textit{Sweetsourcod} is based on the method by Skilling \cite{skilling2004programming} and adapted from \cite{hilbertcurve}.

\section{supplementary data}

\subsection{Active Brownian particles}

\begin{figure*}
\includegraphics[width=0.75\linewidth]{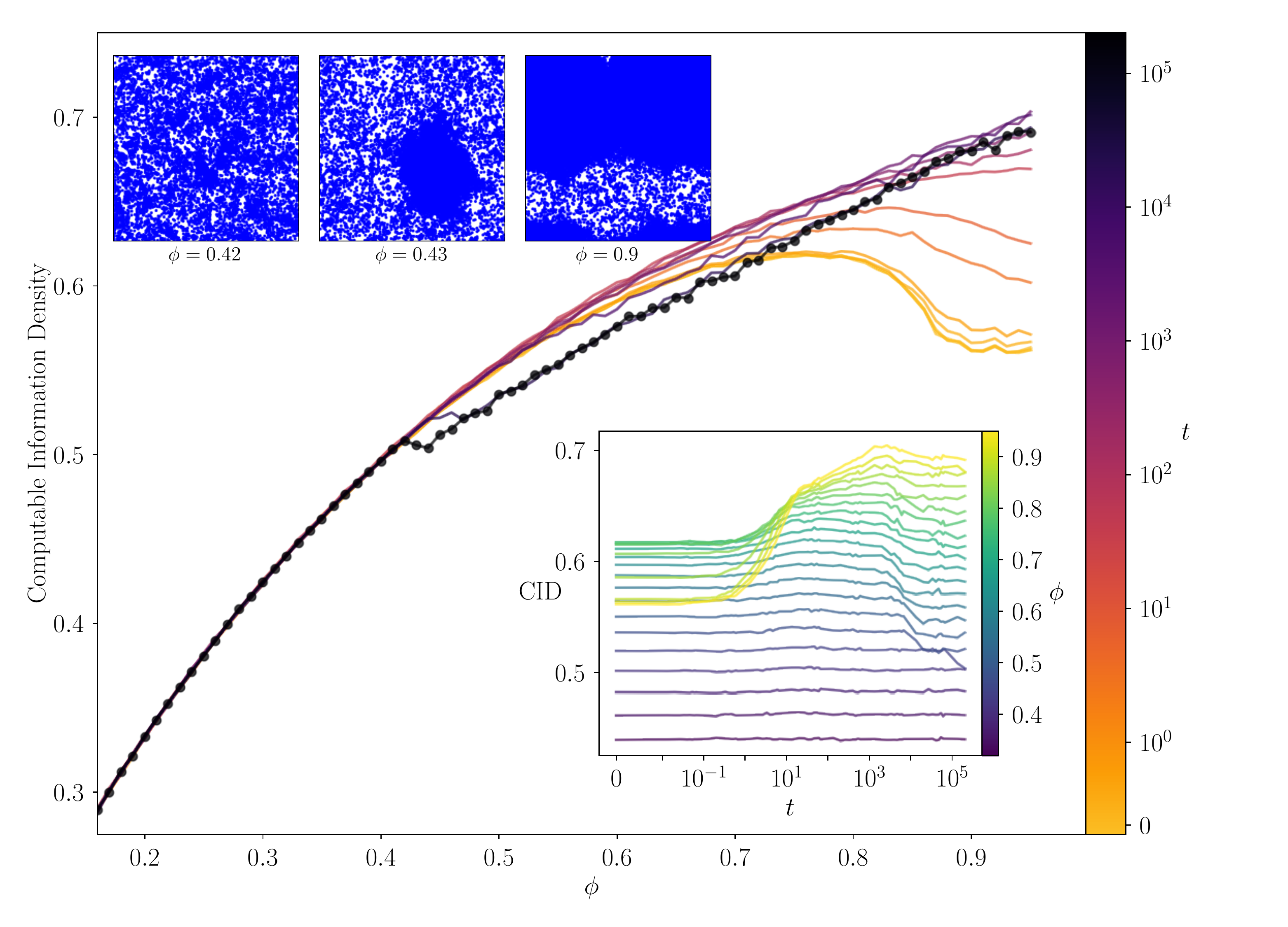}
\protect\caption{Active Brownian Particles:  Systems of 16,384 $\times\rho$  disks with short range repulsion, self-propelling at speed $v_0=0.15$.  Coordinates are quantized (digitized) using a square grid with bin-size approximately $d/\sqrt{5}$. At $\phi \approx 0.43$ the CID drops precipitously, indicating ordering associated with clustering and motility induced phase separation \cite{solon2018gt}. Representative configurations are shown for $\phi = 0.42, \, 0.43,\, 0.90$. For the initial quenched configurations (yellow curve) the flat region for $\phi \gtrapprox 0.88$ corresponds to samples consisting of small grain crystals. The inset shows the time dependence of the CID for different densities.
 \label{fig:swimmers}}
\end{figure*}

In Fig.~\ref{fig:swimmers} we show result analogous to Fig.~5 of the main text for a different particle velocity that shifts the critical point. We prepare the system by depositing $N = 16384 \times \phi$ monodisperse disks in a fixed area for $95$ area fractions in the range $0.01 \leq \phi \leq 0.95$. We  minimize the energy by steepest descent, and then let the system evolve under periodic boundary conditions with velocity $v_0=0.15$, mobility $\mu=1$, rotational diffusion rate $\nu_r=5\times10^{-4}$, and spring constant $k=1$.  We evolve the system according to Eq.~3 (main text) for time $t_{\mathrm{max}}=2 \times 10^5$ and time step $\Delta t = 0.67 \times 10^{-2}$.  Curves are averaged over $6$ independent random initial configurations.

\bibliography{bibliography}